\documentclass[floatfix,citeautoscript,nofootinbib,superscriptaddress,onecolumn,pra]{revtex4-2}
\usepackage{amsbsy}
\usepackage{latexsym,epsfig,graphicx}
\usepackage{dcolumn}
\usepackage{graphicx}
\usepackage{subfigure}
\usepackage{comment}
\usepackage{color}
\usepackage{bm}
\usepackage{mathrsfs}
\usepackage{amsfonts}
\usepackage{amsmath}
\usepackage{amssymb}
\usepackage{xspace}
\usepackage{epstopdf}
\usepackage{tabularx}
\usepackage{longtable}
\usepackage{wasysym}
\usepackage[colorlinks=true, letterpaper=true, pdfstartview=FitV, linkcolor=red, citecolor=blue, urlcolor=blue]{hyperref}
\usepackage[normalem]{ulem}

\pdfoutput=1
\begin{document}
\title{Quadrature-PT symmetry: Classical-to-quantum transition in noise fluctuations}

\author{Wencong Wang}
\affiliation{Guangdong Provincial Key Laboratory of Quantum Engineering and Quantum Materials, School of Physics and Telecommunication Engineering, South China Normal University, Guangzhou 510006, China}
\author{Yanhua Zhai}
\affiliation{Physics Department, Spelman College, Atlanta, Georgia 30314, USA}
\author{Dongmei Liu}
\email{dmliu@scnu.edu.cn}
\affiliation{Guangdong Provincial Key Laboratory of Quantum Engineering and Quantum Materials, School of Physics and Telecommunication Engineering, South China Normal University, Guangzhou 510006, China}
\affiliation{School of Optics, University of Technology, 2000 J St. NW, Washington DC, 20036}
\author{Xiaoshun Jiang}
\email{jxs@nju.edu.cn}
\affiliation{National Laboratory of Solid State Microstructures, College of Engineering and Applied Sciences, Nanjing University, Nanjing 210093, China}
\author{Saeid Vashahri Ghamsari}
\affiliation{Department of Physics, Kennesaw State University, Marietta, Georgia 30060, USA}
\author{Jianming Wen}
\email{jianming.wen@kennesaw.edu}
\affiliation{Department of Physics, Kennesaw State University, Marietta, Georgia 30060, USA}

\begin{abstract} 
While gain-loss-coupled photonic platforms have achieved significant success in studying classical parity-time (PT) symmetry, they encounter challenges in demonstrating pure quantum effects due to incompatible operator transformations and Langevin noise. Here, we present compelling evidence that a non-Hermitian (NH) twin-beam system, undergoing phase-sensitive amplification (PSA) and balanced loss, not only enables observing the usual eigenvalue-associated PT phase transition but also exhibits distinctive features absent in classical NH or Hermitian quantum scenarios, encompassing quadrature PT symmetry, anomalous loss-induced quadrature squeezing, and classical-to-quantum transitions in noise fluctuations. Furthermore, our proposed bipartite open system promises optimal sensing, showcasing an improved signal-to-noise ratio and sensitivity, constrained by quantum Cram\'{e}r-Rao bound or Fisher information. These findings deepen the comprehension of authentic quantum optical PT symmetry involving both gain and loss, addressing contentious issues and illuminating new facets of the subject.

\end{abstract}

\maketitle
%%%%%%%%%%%%%%%%%%%%%%%%%%  body  %%%%%%%%%%%%%%%%%%%%%%%%%%
\section{Introduction}
The Hamiltonian of a physical system is commonly presumed to be Hermitian in order to ensure its real eigenvalues. However, due to the unavoidable interaction with its surroundings, non-Hermitian (NH) effects are prevalent in both classical and quantum domains. Previously, this anti-Hermiticity was regarded as harmful and requiring suppression in measurements and practical applications. In 1998, Bender and Boettcher challenged this perspective by demonstrating that if an NH Hamiltonian commutes with the joint parity-time (PT) operator, its eigenvalues can be entirely real until a phase transition occurs at an exceptional point (EP) \cite{1}. EPs, as unique features of NH physics, are degeneracies where both eigenvalues and eigenvectors merge. These striking properties have stimulated extensive research across various disciplines. In particular, gain-loss-coupled optical systems have become a fertile ground for exploring PT symmetry in both linear and nonlinear cases over the past decade \cite{2,3,4,5,6,7,8,9,10,11,12,13,14,15,16,17}. These studies have unveiled a multitude of peculiar effects and led to innovative technology advancements, including sensing applications.

While classical NH dynamics have been effectively utilized in gain-loss-coupled photonic structures, harnessing them for pure quantum PT effects faces challenges. Previous studies adopted a mean-field approach, treating light as a (semi)classical electromagnetic (EM) field and encapsulating quantum dissipation in an `effective Hamiltonian' \cite{2,3,4,5,6,7,8,9,10,11,12,13,14,15,16,17}. As a result, the illustrated PT phase transitions in single-partite quantum systems are limited to (semi)classical interpretations. Open quantum systems require quantized EM fields and Langevin noise operators to preserve commutation relations \cite{18}. Recent arguments suggest challenges in achieving quantum PT symmetry in photonic systems involving both gain and loss \cite{19}, especially in phase-insensitive amplification (PIA). Adhering to causality \cite{20} and maintaining the original Hilbert space \cite{21} pose difficulties with gain. To circumvent these issues, two approaches have emerged: a passive scheme or an NH subset Hamiltonian within a larger Hermitian system \cite{22,23,24,25,26,27,28}, coupled with postselection measurement. Another research avenue explores anti-PT symmetry \cite{29,30,31,32,33,34,35,36} without requiring gain and loss \cite{37,38}. Despite uncovering some quantum features, the feasibility of gain-loss-based quantum optical PT symmetry, along with fair sampling measurement, remains uncertain.

Here, we address this challenge \cite{19} affirmatively, employing phase-sensitive linear quantum amplification (PSA) on one mode of twin beams generated via nonlinear optical processes like parametric down conversion (PDC) and four-wave mixing (FWM) \cite{39}. This unique bipartite system showcases novel features: PSA's noiseless quantum amplification \cite{40} enables genuine quantum-optical PT experimentation under fair sampling measurement, yielding unique \textit{quadrature PT symmetry} and \textit{anomalous loss-induced quadrature squeezing} and accelerating nonclassical correlation emergence. The system reveals quantum NH properties distinct from classical and Hermitian quantum counterparts, including nontrivial \textit{classical-to-quantum} (C2Q) transitions in PT-manifested quadratures in the dynamical or stationary fashion. Moreover, our theory shows that the PSA-loss interplay may enhance some observables' sensing abilities with improved signal-to-noise ratio (SNR) and precision in some system parameter range, supported by quantum Fisher information (QFI). Furthermore, our findings contribute insights to unresolved issues in previous studies.

\section{Theoretical model}
\begin{figure}[htbp]
\centering\includegraphics[width=0.9\linewidth]{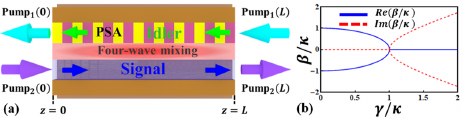}
\caption{(a) Quadrature PT symmetry and loss-induced quadrature squeezing formed by twin beams from a backward-FWM process. The $-z$-idler mode experiences PSA, while the $+z$-signal engages equal loss. (b) Standard PT phase transition exhibited by eigenvalues $\pm\beta$.}
\label{fig:scheme}
\end{figure}

As schematic in Fig.~\ref{fig:scheme}(a), our quadrature-PT structure originates from NH twin-beam generation. This involves driving an $L$-long FWM medium with a vacuum input state, ensuring perfect phase matching and a counter-propagating configuration. The nondegenerate signal-idler modes undergo PSA and loss at rates of $g$ and $\gamma$, respectively. Treating the input pump lasers as nondepleted and classical \cite{39}, the system's evolution along the $\pm z$-direction is thus influenced by the NH Hamiltonian (with SU(1,1) symmetry),
\begin{eqnarray}
H=i\hbar g(a^2_i-a^{\dagger 2}_i)/2-i\hbar\gamma a^{\dagger}_sa_s+\hbar\kappa(a^{\dagger}_ia^{\dagger}_s+a_ia_s),\label{eq:hamiltonian}
\end{eqnarray}
and the signal-idler field operators, $a_s$ and $a_i$, obey the Heisenberg-Langevin equations,
\begin{eqnarray}
\frac{da_i}{dz}=ga^{\dagger}_i+i\kappa a^{\dagger}_s,\; \frac{da_s}{dz}=-\gamma a_s-i\kappa a^{\dagger}_i+f_s,
\label{eq:fieldoperator}
\end{eqnarray}
with $\dagger$ denoting Hermitian conjugate, $\kappa$ the parametric conversion strength, and $f_s$ the Langevin noise operator of zero mean satisfying $\langle f_s(z)f^{\dagger}_s(z')\rangle=2\gamma\delta(z-z')$ and $\langle f^{\dagger}_s(z)f_s(z')\rangle=0$. At a cursory look, dynamics (\ref{eq:fieldoperator}) seems PT-irrelevant, even if $g=\gamma$. However, hidden PT arises when transforming Eqs.~(\ref{eq:fieldoperator}) into the corresponding quadrature-operator evolutions using $q_j=(a^{\dagger}_j+a_j)/2$ and $p_j=i(a^{\dagger}_j-a_j)/2$ ($j=i,s$), where $[q_j,p_j]=i/2$. That is,
\begin{subequations}
\begin{align}
\frac{d}{dz}\begin{bmatrix}
q_i\\
p_s
\end{bmatrix}&=\begin{bmatrix}
g & \kappa\\
-\kappa & -\gamma
\end{bmatrix}\begin{bmatrix}
q_i\\
p_s
\end{bmatrix}+\begin{bmatrix}
0\\
P_s
\end{bmatrix}, \label{eq:quadrature1} \\
\frac{d}{dz}\begin{bmatrix}
p_i\\
q_s
\end{bmatrix}&=\begin{bmatrix}
-g & \kappa\\
-\kappa & -\gamma
\end{bmatrix}\begin{bmatrix}
p_i\\
q_s
\end{bmatrix}+\begin{bmatrix}
0\\
Q_s
\end{bmatrix}, \label{eq:quadrature2} 
\end{align}
\end{subequations}
with the Langevin-noise quadrature operators $P_s=i(f^{\dagger}_s-f_s)/2$ and $Q_s=(f^{\dagger}_s+f_s)/2$. It is crucial to note that selecting a zero-squeezing angle emphasizes the pivotal role of PT, contrasting with the classical quadrature noise in the Hermitian scenario without PSA and loss. The introduced $g$-$\gamma$ pairing here vastly disrupts usual two-mode squeezing generation. Specifically, at $g=\gamma$, the quadrature pair $\{q_i,p_s\}$ (\ref{eq:quadrature1}) undergoes PT-symmetric evolution, displaying a typical phase transition at the EP ($\gamma=\kappa$) where eigen-propagation constants $\pm\beta=\pm\sqrt{\kappa^2-\gamma^2}$ transition from real to imaginary (Fig.~\ref{fig:scheme}(b)). In contrast, the conjugate pair $\{p_i,q_s\}$ (\ref{eq:quadrature2}) experiences loss-induced squeezing, independent of $\{q_i,p_s\}$. These asymmetric dynamics, unreachable to PIA-based NH platforms, also allow for dual opposing quadrature PT symmetry--$\{q_i,p_s\}$ for active while $\{p_i,q_s\}$ for passive when $\gamma=0$ \cite{40}. To differentiate between these two scenarios, we designate the case where $g=\gamma\neq0$ as type-I, and the case where $g\neq0,\gamma=0$ as type-II. In this study, our focus is solely on type-I quadrature PT symmetry. The general solutions to Eqs.~(\ref{eq:quadrature1}) and (\ref{eq:quadrature2}) are 
\begin{subequations}
\begin{align}
\begin{bmatrix}
q_i(0)\\
p_s(L)
\end{bmatrix}&=\sec(\beta L-\epsilon)\begin{bmatrix}
\cos\epsilon & -\sin(\beta L)\\
-\sin(\beta L) & \cos\epsilon
\end{bmatrix}\begin{bmatrix}
q_i(L)\\
p_s(0)
\end{bmatrix}\nonumber\\
&+\sec(\beta L-\epsilon)\int_0^LdzP_s(z)\begin{bmatrix}
-\sin[\beta(L-z)]\\
\cos(\beta z-\epsilon)
\end{bmatrix}, \label{eq:solution1} \\
\begin{bmatrix}
p_i(0)\\
q_s(L)
\end{bmatrix}&=\sec(\kappa L)\begin{bmatrix}
e^{\gamma L} & -\sin(\kappa L)\\
-\sin(\kappa L) & e^{-\gamma L}
\end{bmatrix}\begin{bmatrix}
p_i(L)\\
q_s(0)
\end{bmatrix}\nonumber\\
&+\sec(\kappa L)\int_0^LdzQ_s(z)\begin{bmatrix}
-e^{\gamma z}\sin[\kappa(L-z)]\\
e^{\gamma(z-L)}\cos(\kappa z)
\end{bmatrix}, \label{eq:solution2} 
\end{align}
\end{subequations}
with a PT-induced phase $\epsilon=\arctan(\gamma/\beta)$. The solutions (\ref{eq:solution1}) assure that any observation of $\{q_i(0),p_s(L)\}$ is governed by quadrature PT. Additionally, owing to the inseparability of the real signal and Langevin noise, measurements on these quadratures will inevitably be disturbed by the Langevin noise. Further discussions on these results, including the preservation of their commutation relationships, can be found in the Supplementary Information (SI).

\section{Homodyne detection}
To unveil the underlying physics behind the results (\ref{eq:solution1}) and (\ref{eq:solution2}), a direct way is to analyze the noise characteristics of these single-mode quadratures, contrasting them with the correspding Hermitian case and vacuum noise. This leads us to examine four variances: $\langle\Delta q_j^2\rangle=\langle q^2_j\rangle-\langle q_j\rangle^2$ and $\langle\Delta p^2_j\rangle=\langle p^2_j\rangle-\langle p_j\rangle^2$. After some algebra (SI), we have
\begin{subequations}
\begin{align}
\langle\Delta q^2_i(0)\rangle&=\frac{h(L)-2\sin^2\epsilon-\sec\epsilon\cos(2\beta L-\epsilon)}{8\cos^2(\beta L-\epsilon)}, \label{eq:qivariance} \\
\langle\Delta p^2_s(L)\rangle&=\frac{h(L)-\cos(2\beta L)+\tan\epsilon\sin(2\beta L-2\epsilon)}{8\cos^2(\beta L-\epsilon)}, \label{eq:psvariance}\\
\langle\Delta q^2_s(L)\rangle&=\frac{2+\cos^2\varphi e^{-2\gamma L}-\cos\varphi\cos(2\kappa L+\varphi)}{8\cos^2(\kappa L)}, \label{eq:qsvariance} \\
\langle\Delta p^2_i(0)\rangle&=\frac{(2+\cos^2\varphi)e^{2\gamma L}-\cos\varphi\cos(2\kappa L-\varphi)}{8\cos^2(\kappa L)}, \label{eq:pivariance}
\end{align}
\end{subequations}
with $h(L)=3+2\gamma L$ and $\varphi=\tan^{-1}(\gamma/\kappa)$. Notably, aside from the typical $\pm\beta$-bifurcation (Fig.~\ref{fig:scheme}(b)), PT introduces significant differences in variances (\ref{eq:qivariance}) and (\ref{eq:psvariance}) compared to (\ref{eq:qsvariance}) and (\ref{eq:pivariance}), as reflected by the argument $\beta L$. 

\begin{figure}[htbp]
\centering\includegraphics[width=0.8\linewidth]{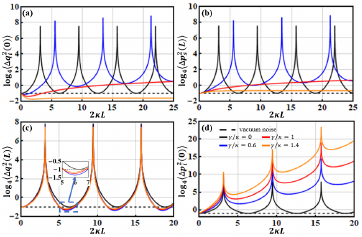}
\caption{Asymmetric PSA amplification gives rise to distinctive noise fluctuations in single-mode PT-quadratures $\{q_i(0),p_s(L)\}$, resulting in a unique C2Q transition in $\langle\Delta q^2_i(0)\rangle$ (a) and a classical transition in $\langle\Delta p^2_s(L)\rangle$ (b). Similarly, single-mode non-PT-quadratures $\{q_s(L),p_i(0)\}$ exhibit anomalous loss-induced squeezing in $\langle\Delta q^2_s(L)\rangle$ (c) and anti-squeezing in $\langle\Delta p^2_i(0)\rangle$ (d), diverging from the Hermitian TMSV (black solid) and the vacuum noise (dashed black).}
\label{fig:singlevariance}
\end{figure}

Numerically illustrating these variances in Fig.~\ref{fig:singlevariance} for various $\gamma/\kappa$ values, we compare them with the Hermitian case (solid black lines) and vacuum noise level (dashed black). Figures~\ref{fig:singlevariance}(a) and (b) exhibit a number of unusual noise behaviors due to nontrivial interference between parametric conversion and PSA/loss, inaccessible in PIA-based PT or traditional two-mode vacuum squeezing (TMVS) platforms. In the PT-phase intact region ($\gamma/\kappa<1$), the quantum observables $q_i(0)$ and $p_s(L)$ are overall anti-squeezed, with noises (blue) fluctuating classically at a period $\approx2\pi\kappa/\beta$, deviating dramatically from the $2\pi$-period of the TMSV. Conversely, in the PT-phase broken regime ($\gamma/\kappa>1$), both $\langle\Delta q^2_i(0)\rangle$ and $\langle\Delta p^2_s(L)\rangle$ (orange) cease to oscillate logarithmically and are upper-bounded by their respective EP curves (red). Additionally, $\langle\Delta q^2_i(0)\rangle$ can be lower than the vacuum noise within a dimensionless propagation length $2\kappa L$, indicating the speedup of quantum squeezing emergence in comparison with the traditional TMVS case; while $\langle\Delta p^2_s(L)\rangle$ displays classical noise exponential amplification. Moreover, as $\gamma/\kappa$ increases, $q_i(0)$ exhibits greater quantum squeezing, and $p_s(L)$ experiences less classical noise over a longer $2\kappa L$. Besides the standard PT phase transition at $\pm\beta$, the $q_i(0)$-noise (Fig.~\ref{fig:singlevariance}(a)) further manifests a radical classical-to-quantum (C2Q) transition for a fixed length by simply increasing $\gamma$. The boundary of the C2Q transition hinges on whether the length is smaller or greater than the intersection point set by the EP curve (red) with the vacuum noise (dashed black) in Fig.~\ref{fig:singlevariance}(a). Specifically, before reaching this intersection length, the EP-variance curve acts as the boundary, distinguishing classical noise from quantum squeezed noise as $\gamma$ gradually increases. However, beyond this intersection point, the EP-variance curve ceases to be the partition line separating the two realms of noise behaviors, and the precise transition boundary deviates from the EP-curve, becoming contingent on the length of interest. This pure quantum NH effect, unattainable in single-partite quantum NH schemes, offers a unique tool for probing the boundary between classical and quantum worlds--a fundamental challenge in quantum mechanics. On the other hand, despite loss and Langevin noise, $\{p_i(0),q_s(L)\}$ variances (Figs.~\ref{fig:singlevariance}(c) and (d)) dramatically departure from ideal TMSV, signifying anti-squeezing and loss-induced quadrature squeezing at some $2\kappa L$ (e.g., dashed blue rectangle in $\langle\Delta q^2_s(L)\rangle$), an abnormal effect impossible in PIA-based PT settings. The appearance of loss-induced quadrature squeezing is counter-intuitive to conventional wisdom, where loss is typically considered to degrade quantum squeezing effect. In this context, the \textit{loss} terms in Eq.~(\ref{eq:quadrature2}) unexpectedly lead to the manifestation of noise reduction below the vacuum noise level in the observable $q_s(L)$ at an earlier stage compared to the vacuum-input TMSV case (Figs.~\ref{fig:singlevariance}(c)). 

\begin{figure}[htbp]
\centering\includegraphics[width=0.9\linewidth]{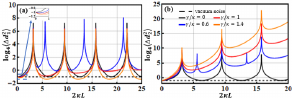}
\caption{PT-symmetric noise fluctuations in two-mode quadratures: $\langle\Delta d^2_1\rangle$ (a) exhibits a C2Q transition with dual fluctuating periods before the standard PT-phase transition of $\pm\beta$, transforming into a single oscillating period afterward. Conversely, $\langle\Delta d^2_2\rangle$ (b) only shows classical fluctuations, transitioning from dual periods to a single period. The black solid and dashed curves represent the ideal TMSV and vacuum noise for comparison.}
\label{fig:twovariance}
\end{figure}

The analysis reveals that our PSA-loss twin-beam system enables simultaneous observation of two distinctive transitions based on eigenvalues and noise characteristics in single-mode quadrature measurements. This prompts exploration of its existence in two-mode quadrature homodyne detection. We define the following two-mode quadratures as $d_1=[q_i(0)+q_s(L)]/\sqrt{2}$ and $d_2=[p_i(0)+p_s(L)]/\sqrt{2}$, satisfying $[d_1,d_2]=i/2$. Simple calculations show that for the vacuum input, their variances are the sum of the single-mode variances (\ref{eq:qivariance})--(\ref{eq:pivariance}),
\begin{eqnarray}
\langle\Delta d^2_1\rangle=\frac{\langle\Delta q^2_i(0)\rangle+\langle\Delta q^2_s(L)\rangle}{2},
\langle\Delta d^2_2\rangle=\frac{\langle\Delta p^2_s(L)\rangle+\langle\Delta p^2_i(0)\rangle}{2}.
\label{eq:d2var}
\end{eqnarray}
Thanks to these inter-relationships, the noise properties of the observables $d_1$ and $d_2$ resemble those of $\{q_i(0),p_s(L),q_s(L),p_i(0)\}$ but with noticeable differences. Similar to $q_i(0)$, dual transitions occur in $d_1$, with its EP-variance curve acting as the lower boundary between classical and quantum noise distributions when $\gamma/\kappa\geq1$, as depicted in Fig.~\ref{fig:twovariance}(a). Before the PT-phase transition at $\pm\beta$, $\langle\Delta d^2_1\rangle$ follows a two-period classical noise fluctuation superimposed on the vacuum noise, except for a very short distance range highlighted in the inset. However, after the phase transition, it transforms into a single-period fluctuation and demonstrates quantum squeezing for certain values of $2\kappa L$. Moreover, as $\gamma/\kappa$ increases, this nonclassical effect persists for larger $2\kappa L$ values. The abrupt change in noise nature establishes $d_1$ as another quantum observable to research the C2Q transition. Regarding $d_2$, as shown in Fig.~\ref{fig:twovariance}(b), its noise is consistently anti-squeezed and gives interleaved double periodic fluctuations before the PT phase transition at $\pm\beta$, and a single oscillating period thereafter.

\section{RISM}
\begin{figure}[htbp]
\centering\includegraphics[width=0.9\linewidth]{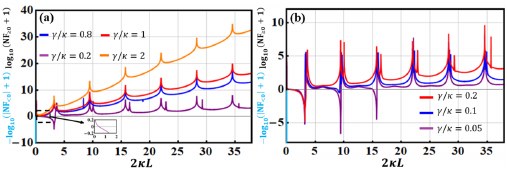}
\caption{(a) NF with PT manifestation across the EP. (b) Representative examples of NF, indicating the possibility of observing relative-intensity squeezing after the C2Q transition in the PT-phase unbroken region.}
\label{fig:NF}
\end{figure}

Other than homodyne detection, we can utilize the \textit{relative intensity squeezing measurement} (RISM) to explore the proposed quadrature PT effect. Traditionally, this method involves measuring the shot-noise of one beam and subtracting it from the other, resulting in a lower-noise differential measurement of a signal of interest. The process utilizes the relative-intensity operator, $N_i(0)-N_s(L)=a^{\dagger}_i(0)a_i(0)-a^{\dagger}_s(L)a_s(L)$, with the noise figure (NF) as a parameter characterizing squeezing. The NF \cite{41} is determined by the relative-intensity variance and assumes a mathematical form, $\mathrm{NF}\equiv\mathrm{Var}[N_i(0)-N_s(L)]/\langle N_i(0)\rangle+\langle N_s(L)\rangle$, with the average photon numbers of $\langle N_i(0)\rangle=\langle q^2_i(0)\rangle+\langle p^2_i(0)\rangle-1/2$ and $\langle N_s(L)\rangle=\langle q^2_s(L)\rangle+\langle p^2_s(L)\rangle-1/2$ (SI).
The NF measurement reveals distinct characteristics from the discussed quadrature variances. In Fig.~\ref{fig:NF}(a), in the PT-phase broken regime ($\gamma/\kappa\geq1$), $\mathrm{NF}_{\geq0}+1$ shows classical noise periodic fluctuation in logarithm, with higher $\gamma/\kappa$ values leading to noisier NF. Contrarily, before PT symmetry breaks down ($\gamma/\kappa<1$), the NF behaves complex. The NF curve at the EP no longer serves as the partition to separate classical noise fluctuation from quantumly squeezed fluctuation. Our simulations indicate that the NF curve at $\gamma/\kappa\approx0.52$ acts as the C2Q boundary. Remarkably, in the region where $\gamma/\kappa<0.52$, $-\lg(|\mathrm{NF}_{<0}|+1)$ can become negative for some $2\kappa L$, implying the emergence of relative-intensity squeezing. In Fig.~\ref{fig:NF}(b), representative examples with different $\gamma/\kappa$ values illustrate how smaller $\gamma/\kappa$ values lead to larger relative-intensity squeezing for larger $2\kappa L$. Overall, the observable NF offers a different perspective on the noise behavior of the system and the C2Q transition, making the RISM a valuable complementary measure to the previously mentioned homodyne measurements.

Before proceeding, a few key remarks: (i) Utilizing PSA instead of PIA as the gain mechanism in a \textit{bipartite} system achieves true quantum-optical PT symmetry without postselection. (ii) Besides the regular PT phase transition of eigenvalues, observable purely quantum NH effects, such as the three C2Q transitions in noise fluctuations reported here, are inaccessible to classical NH and Hermitian quantum counterparts. (iii) Applying PSA in the continuous variable (CV) basis, as done here, preserves the quantum noncloning theorem and avoids expanding the Hilbert space. (iv) The single-mode nature of the signal and idler ensures compliance with the causality principle in our bipartite system. (v) Squeezing leads to Einstein-Podolsky-Rosen-type correlations in photon-number fluctuations between the signal and idler, even in the presence of Langevin noise-induced thermalization.

\section{Quantum sensing}

As a discipline dedicated to practical applications, quantum sensing \cite{42,43,44} leverages quantum properties, effects, or systems for high-resolution and super-sensitive measurements of physical parameters compared to classical frameworks. Quantum squeezing, a vital nonclassical resource for ultra-precise estimations, has recently found implementation in the Laser Interferometer Gravitational-Wave Observatory (LIGO) for gravitational wave detection. However, inherent propagation loss often degrades available squeezing and compromises promised sensitivity. In recent NH studies, the abrupt eigenspectral change near EP has been capitalized for enhanced sensing in classical settings \cite{45,46,47,48,49}. Extending this to the quantum level is challenging due to Langevin noise \cite{50}. Common strategies to overcome this noise involve ideal anti-PT systems or passive NH schemes along with postselection measurements \cite{26,27,28,37}. In contrast, our quadrature-PT system promises improved SNR and sensitivity in quantum sensing under fair sampling measure. As an example, we utilize single-mode quadratures to illustrate our sensing protocol and analyze it by estimating $\kappa$ and comparing the achievable precision with the quantum Cram\'{e}r-Rao bound, set by the quantum Fisher information of the quantum state.

Assume the initial preparation of the two bosonic modes in a coherent state $|\Phi_0\rangle=|\alpha_i,\alpha_s\rangle$. An optimal homodyne detection is then conducted on, say, $q_i(0)$. Using Eq.~(\ref{eq:solution1}), we can obtain its mean value and variance,
\begin{eqnarray}
&\langle q_i(0)\rangle=\frac{\cos\epsilon\langle q_i(L)\rangle-\sin(\beta L)\langle p_s(0)\rangle}{\cos(\beta L-\epsilon)},\label{eq:qi}\\
&\langle\Delta q^2_i(0)\rangle=\frac{3+w[2\gamma L-\tan\epsilon\sin(2\beta L)]-\cos(2\beta L)-2\sin^2\epsilon}{8\cos^2(\beta L-\epsilon)},\label{eq:qivar}
\end{eqnarray}
where $w=2n_{\mathrm{th}}+1$ with $n_{\mathrm{th}}$ the average thermal boson number. Note that Langevin noise shifts the peaks of $\langle\Delta q^2_i(0)\rangle$ away from the troughs of $\langle q_i(0)\rangle$, preventing their overlap. To ease subsequent derivations, we assume $\alpha_i=i\alpha^*_s\equiv\sqrt{2}\alpha e^{i\pi/4}$. Estimation precision relies on measuring the rapid change of $\langle q_i(0)\rangle$ caused by a tiny perturbation $\delta\kappa$ on a preset $\kappa$. This suggests probing the system's response to $\langle q_i(0)\rangle$ for a small variation $\delta\kappa$ around $\kappa$. We define the `susceptibility $\chi^{q_i(0)}_{\kappa}\equiv\partial\langle q_i(0)\rangle/\partial\kappa$' to capture this system response,
\begin{eqnarray}
&\frac{\alpha\{2\beta L[\sin(\beta L-\epsilon)-1]+\sin\epsilon[2\sin(\beta L)+\cos(\beta L-\epsilon)-\cos\epsilon]\}}{2\beta\cos^2(\beta L-\epsilon)}.\label{eq:chi}
\end{eqnarray}
As $\kappa\rightarrow\gamma$ or $\beta\rightarrow0$, $\chi^{q_i(0)}_{\kappa}\rightarrow-\alpha L(3+\kappa^2L^2)/[3(1+\kappa L)^2]$, presenting a curvatureless profile (Fig. S2 in SI). This implies that the system becomes less sensitive to perturbations at the EP. Coversely, $\kappa$-estimation is jointly determined by the variance and susceptibility via the relation $\Delta\kappa^2_{q_i(0)}=\langle\Delta q^2_i(0)\rangle/[\chi^{q_i(0)}_{\kappa}]^2$. The variance inverse, $\Delta\kappa^{-2}_{q_i(0)}$, dictates the SNR and its exact expression is given in SI.

\begin{figure}[htbp]
\centering\includegraphics[width=0.9\linewidth]{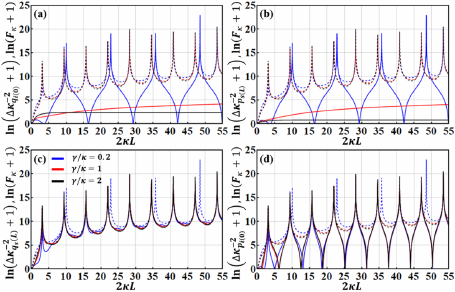}
\caption{Quantum sensing for $\kappa$-estimation by assessing the ratios of $\Delta\kappa^{-2}_{q_i(0)}$ (solid) (a), $\Delta\kappa^{-2}_{p_s(L)}$ (solid) (b), $\Delta\kappa^{-2}_{q_s(L)}$ (solid) (c), and $\Delta\kappa^{-2}_{p_i(0)}$ (solid) (d) to the quantum Fisher information $F_{\kappa}$ (dotted) with parameters $\{\alpha=2,\gamma=0.2,\kappa=1\}$ (blue), $\{2,1,1\}$ (red), and $\{2,2,1\}$ (orange).}
\label{fig:sensing}
\end{figure}

The protocol performance is then to compare $\Delta\kappa^{-2}_{q_i(0)}$ with the quantum Fisher information, $F_{\kappa}$, which sets the ultimate precision or the lower quantum Cram\'{e}r-Rao bound for optimal measurements, $F_{\kappa}\geqslant\Delta\kappa^{-2}_{q_i(0)}$ (Fig.~\ref{fig:sensing}). $F_{\kappa}$ is deduced from $F_{\kappa}=(d\mu_{\mathrm{out}}/d\kappa)^{T}V^{-1}_{\mathrm{out}}(d\mu_{\mathrm{out}}/d\kappa)$, where the amplitude vector $\mu_{\mathrm{out}}$ and variance matrix $V_{\mathrm{out}}$ are computed in the quadrature basis (see SI for detail).
%via $\mu_j=\langle\hat{\nu}_j\rangle$ and $V_{j,k}=\langle\hat{\nu}_j\hat{\nu}_k+\hat{\nu}_k\hat{\nu}_j\rangle/2-\langle\hat{\nu}_j\rangle\langle\hat{\nu}_k\rangle$, for $1\leq j,k\leq2$, with the column vector $\hat{\nu}=(q_i(0),q_s(L),p_i(0),p_s(L))^T$. 
Similar sensing evaluations for the remaining quadratures can be performed, as outlined in the SI. Comparing $\Delta\kappa^{-2}_{q_i(0)}$ and $\Delta\kappa^{-2}_{p_s(L)}$ to $F_{\kappa}$ (Figs.~\ref{fig:sensing}(a) and (b)) indicates that, in the current setup, $q_i(0)$ and $p_s(L)$ yield optimal classical sensitivity at turning points of $L=(2n-1/2)\pi/\beta$, with positive integers $n$, in the PT-phase unbroken regime. In contrast, for the non-PT quadrature pair (Figs.~\ref{fig:sensing}(c) and (d)), both optimal classical and quantum sensing are doable despite loss and Langevin noise. Precisely, optimal classical sensing is attainable at locations where $L=(2n-1/2)\pi/\kappa$ using either $q_s(L)$ or $p_i(0)$. In stark contrast, $q_s(L)$ further permits optimal quantum sensing at $L=n\pi/\kappa$, with larger $\gamma/\kappa$ values corresponding to enhanced sensitivity and SNR.

\section{Conclusion}

In summary, contrary to previous perspectives, the proposed bipartite NH system stands out as a robust platform for exploring genuine quantum optical PT symmetry within the CV basis. It features distinct attributes unattainable in discrete-variable frameworks or PIA-based NH systems. Notable among these are quadrature PT symmetry, the C2Q transition in noise fluctuations, and loss-induced quadrature squeezing. Any purported authentic quantum PT effect must exhibit additional transitions beyond usual eigenvalue-based PT phase shifts to substantiate its quantum nature, avoiding (semi) classical interpretations. The PT quadrature pair, facilitated by an expanded Hilbert space from amplified Langevin noise, offers optimal classical sensing, while the non-PT quadrature pair accommodates both optical classical and quantum sensing. Importantly, while optics serves as an illustrative example, this model holds applicability across various platforms, including superconducting circuits. Moreover, our work pioneers a novel avenue for exploring elusive nontrivial C2Q transitions, harnessing NH physics and symmetry in a groundbreaking manner.

\section*{Funding} National Science Foundation (NSF) (2329027, 1806519, and EFMA-1741693); The National Key R$\&$D Program of China (2021YFA1400803); GuangDong Basic and Applied Basic Research Foundation (2022A1515140139)

\section*{Acknowledgments} We thank Y. Huang, F. Narducci, M. Xiao, J. Jing, Z. Zhou, I. Novikova, Q. Gu, S.-W. Huang, and L. Jiang for enlightening discussions.

\section*{Disclosures} The authors declare no conflicts of interest.

\section*{Supplemental document}
See Supplement 1 for supporting content.

% Bibliography

\end{document}

\title{Supplementary Materials for Quadrature-PT symmetry: Classical-to-quantum transition in noise fluctuations}

\begin{abstract}
\end{abstract}

\maketitle

\section{Further discussion on type-I quadrature PT symmetry}

\subsection{Further remarks on type-I quadrature PT symmetry}
In our previous work \cite{1}, we theoretically demonstrated that a forward parametric nonlinear optical process can lead to anti-PT symmetry while a backward parametric nonlinear optical process can result in PT symmetry. Therefore, we are particularly interested in investigating a backward nonlinear parametric optical process, like backward four-wave mixing, as depicted in Fig. 1(a) in the main text. In this process, the two counter-propagating parametric modes, idler and signal, experience balanced phase-sensitive linear quantum amplification (PSA) and attenuation within the medium of length $L$. The evolution of the paired idler and signal field operators in this type-I PSA-loss twin-beam system is effectively governed by the non-Hermitian (NH) Hamiltonian $H$ (Eq. (1)) as provided in the main text. By using $H$, one can readily derive the Heisenberg equations of the idler-signal field operators,
\begin{subequations}
\begin{align}
i\hslash\frac{\partial a_{i}}{\partial\left( {- z} \right)} = \left\lbrack {a_{i},H} \right\rbrack, \label{eq:Heisenbergequation1} \\
i\hslash\frac{\partial a_{s}}{\partial z} = \left\lbrack {a_{s},H} \right\rbrack, \label{eq:Heisenbergequation2}
\end{align}
\end{subequations}
Thanks to noiseless quantum linear amplification enabled by the PSA \cite{2,3,4,5}, the idler-mode dynamics remains free from additive noise, preserving its commutation relation throughout the entire process. However, this doesn’t hold true for the lossy signal propagation. To restore the commutation relation, it is necessary to introduce quantum Langevin noise in the Heisenberg equation of the signal field operator (\ref{eq:Heisenbergequation2}). This leads us to the coupled Heisenberg-Langevin equations (2) described in the main text for the system of interest. To gain insight into the underlying physics, we transform the creation and annihilation operators, $a_{j}^{\dagger}$ and $a_{j}$ ($j=i,s$), into the corresponding quadrature operators, $X_{1j} = \left( {a_{j}^{\dagger}e^{i\theta} + a_{j}e^{- i\theta}} \right)/2$ and $X_{2j} = i\left( {a_{j}^{\dagger}e^{i\theta} - a_{j}e^{- i\theta}} \right)/2$, where $\theta$ is the squeezing angle. In this work, for simplicity, we set $\theta = 0^{{^\circ}}$ and defer the general case for further investigation. With $\theta = 0^{{^\circ}}$, ${X_{1j} = q}_{j} = \left( {a_{j}^{\dagger} + a_{j}} \right)/2$ and $X_{2j} = p_{j} = i\left( {a_{j}^{\dagger} - a_{j}} \right)/2$. By substituting $a_{j}^{\dagger}$ and $a_{j}$ with $q_{j}$ and $p_{j}$, Eqs. (2) are transformed into two sets of coupled-quadrature equations (3a) and (3b) given in the main text.

Here, it is noteworthy that in conventional quantum squeezing research \cite{2,3,4,5}, the predominant use of PSA is to \textit{statically} amplify a specific quadrature, often overlooking the \textit{dynamic} amplification process proposed in our work. On the other hand, we are aware that recent studies \cite{6,7} have delved into applying PSA to the evolution of both modes in two quantum squeezing generations within two coupled nonlinear optical resonators. It is crucial to note that these systems \cite{6,7} differ fundamentally from ours, relying on the pre-existence of squeezing in two nonlinear optical subsystems for evolution before a beam splitter-like interchange interplay occurs. In contrast, our system operates without the need for initial squeezing generation in the two-mode parametric conversion process. Importantly, the fundamental disparity between these systems \cite{6,7} and ours lies in the impossibility for the former to exhibit quadrature PT symmetry and the associated C2Q transition effect—a distinguishing feature of our approach.

When $g=\gamma$ in Eq. (3a), the effective NH Hamiltonian matrix governing the dynamics of the quadrature-PT pair $(q_i,p_s)$ takes the form:

\begin{subequations}
\begin{align}
H_{({q_{i},p_{s}})} = \begin{bmatrix}
{i\gamma} & {i\kappa} \\
{- i\kappa} & {- i\gamma}
\end{bmatrix}. \label{eq:effect_H_qips}
\end{align}
\end{subequations}
The PT symmetry of $H_{({q_{i},p_{s}})}$ is apparent, as it adheres to the condition of $PTH_{({q_{i},p_{s}})}=H_{({q_{i},p_{s}})}PT$, where $P$ signifies the parity operator $\begin{bmatrix}0 & 1 \\1 & 0\end{bmatrix}$ and $T$ denotes the complex conjugation. Verification of this joint symmetry can be established by applying $PTH_{({q_{i},p_{s}})}$ and $H_{({q_{i},p_{s}})}PT$ to the eigenvectors $\psi_{\pm}=\begin{bmatrix}\frac{\pm i\sqrt{\kappa^2-\gamma^2}-\gamma}{\kappa}\\1\end{bmatrix}$, followed by confirming their equality. Thus, we refer to $(q_i,p_s)$ as the PT-quadrature pair, subject to the governance of PT symmetry. This stands in stark contrast to the effective NH Hamiltonian matrix for the non-PT-symmetric conjugate quadrature pair $(p_i,q_s)$ given by Eq. (3b) in the main text,
\begin{subequations}
\begin{align}
H_{({p_{i},q_{s}})} = \begin{bmatrix}
{- i\gamma} & {i\kappa} \\
{- i\kappa} & {- i\gamma}
\end{bmatrix}. \label{eq:effect_H_piqs}
\end{align}
\end{subequations}
The two eigenvalues of $H_{({q_{i},p_{s}})}$ (\ref{eq:effect_H_qips}) are $\pm\beta=\pm\sqrt{\kappa^2-\gamma^2}$, which gives rise to a regular PT phase transition from real to imaginary. Specifically, when $\gamma/\kappa<1$, the quadrature PT-phase unbroken regime is established, and $\pm\beta$ are real. Conversely, for $
\gamma/\kappa>1$, quadrature PT symmetry spontaneously breaks down, and $\pm\beta$ become purely imaginary. This quadrature PT phase transition occurs at the singular or exceptional point (EP), $\gamma/\kappa=1$, where both eigenvalues and eigenvectors coalescence. Moreover, this PT phase transition significantly alters the properties of the quadrature operators, leading to striking C2Q transitions in noise fluctuations, which are impossible in classical NH or Hermitian quantum counterparts. The combination of $\pm\beta$-spectral bifurcation and C2Q transitions constitutes the proposed "quadrature PT symmetry."

As mentioned in the main text, the asymmetric dynamics facilitated by the PSA-loss twin-beam system can also enable \textit{dual opposing quadrature PT symmetry}, specifically, $(q_i,p_s)$ for active quadrature PT symmetry and $(p_i,q_s)$ for passive quadrature PT symmetry when $\gamma=0$. To differentiate between these two scenarios, we categorize the case when $g=\gamma\neq0$ as type-I and the case where $g\neq0,\gamma=0$ as type-II. In this study, our emphasis is exclusively on type-I quadrature PT symmetry. On the other hand, the main text emphasizes that our work’s ultimate novelty lies not only in identifying a system capable of observing genuine quantum optical PT symmetry under fair sampling measurement but also unveiling an extraordinary phenomenon previously undiscovered. This phenomenon involves the PT-quadrature observable, which allows probing the nontrivial dynamical or stationary C2Q transition alongside the PT phase transition, with a well-defined, adjustable boundary that separates classical noise distribution from quantum squeezed noise by varying the NH parameter $\gamma$ and the dimensionless length $2\kappa L$. In other words, the C2Q transition boundary can coincide with the variance curve at the EP or substantially deviate from it. 

In the literature, there exist proposals \cite{8,9} employing massive quantum oscillators like cooled cavity optomechanical structures to study the quantum-to-classical (Q2C) boundary by monitoring the decoherence of the oscillator’s quantum state while manipulating system parameters. However, even if these proposals were viable in the lab, they face an inescapable challenge--determining the exact transition boundary becomes extremely difficult. In contrast, our system does not encounter such difficulties. Moreover, our method aims to measure the expectation value of a quantum observable, while existing protocols concentrate on studying the system’s state. This fundamental difference sets our work apart from all others. Mostly crucially, our work pioneers a novel approach for investigating the C2Q transition, wherein the system begins in the classical domain. Through the intricate interplay of PSA and the propagation loss, we demonstrate the capability to transition the system into the quantum regime by leveraging non-Hermiticity and symmetry. Distinct from the conventional Q2C transition, this groundbreaking achievement marks the first discovery of this kind in the field.

Before delving into further discussions, it is essential to underscore that for any claimed PT symmetry to be a genuine quantum effect, the NH system must demonstrate an additional nonclassical transition effect, in addition to the typical eigenvalue-based PT phase transition, which can happen in both NH classical and quantum platforms. In the continuous-variable (CV) case, an example of such an effect is the disclosed \textit{C2Q transitions in noise fluctuations} observed in the type-I PSA-loss-coupled twin-beam system. In the discrete-variable (DV) case, an example is the \textit{exceptional entanglement transition} demonstrated in a dissipative spin-boson superconducting system \cite{10,11}. Both the C2Q transition and exceptional entanglement transition lack classical counterparts and are not observable in their corresponding Hermitian settings. A common feature among these new studies is that the analyzed systems are all \textit{bipartite}, a prerequisite for forming nonclassical correlations and entanglements. This, in turn, suggests that previously reported ‘quantum’ PT phase transitions based on \textit{single-partite} systems are open to (semi) classical interpretations. On the other hand, in quantum mechanics, the observation of a physical observable is related to the eigenvalues and eigenstates. As a result, the associated PT phase transition will naturally transform the measurement outcomes, thereby enriching and manifesting effects that cannot be simulated in classical NH or Hermitian quantum systems.

In addition, it is noteworthy that the implementation of PSA accelerates the transition of $q_{i}(0)$ and $q_{s}(L)$ into the quantum realm beforehand. This revelation offers fresh insights into safeguarding CV qubits from decoherence during lossy transmission--a conundrum that has impeded the advancement of diverse CV-based quantum technologies \cite{12}. A more in-depth exploration of this matter exceeds the current scope of this study and will be pursued in future endeavors \cite{13}.

\subsection{Verification of the commutation relations}
Using the solutions (4a) and (4b) presented in the main text, it is straightforward to show that they consistently satisfy the required commutation relations:
\begin{subequations}
\begin{align}
\left\lbrack {q_{i}(0),p_{i}(0)} \right\rbrack &= \frac{e^{\gamma L}cos\epsilon}{cos\left( {\beta L - \epsilon} \right)cos\left( {\kappa L} \right)}\left\lbrack {q_{i}(L),p_{i}(L)} \right\rbrack\nonumber\\ 
&+ \frac{sin(\beta L)sin(\kappa L)}{cos\left( {\beta L - \epsilon} \right)cos\left( {\kappa L} \right)}\left\lbrack {p_{s}(0),q_{s}(0)} \right\rbrack\nonumber\\ 
&+ {\int_{0}^{L}{dz}}\frac{e^{\gamma z}sin\left( \beta\left( {L - z} \right) \right)sin\left( \kappa\left( {L - z} \right) \right)}{cos\left( {\beta L - \epsilon} \right)cos\left( {\kappa L} \right)}\left\lbrack {P_{s},Q_{s}} \right\rbrack = \frac{i}{2}, \label{eq:commutation relations_qipi} \\
\left\lbrack {q_{s}(L),p_{s}(L)} \right\rbrack &= \frac{sin(\beta L)sin(\kappa L)}{cos\left( {\beta L - \epsilon} \right)cos\left( {\kappa L} \right)}\left\lbrack {p_{i}(L),q_{i}(L)} \right\rbrack\nonumber\\ 
&+ \frac{e^{- \gamma L}{\cos\epsilon}}{cos\left( {\beta L - \epsilon} \right)cos\left( {\kappa L} \right)}\left\lbrack {q_{s}(0),p_{s}(0)} \right\rbrack\nonumber\\ 
&+ {\int_{0}^{L}{dz}}\frac{e^{\gamma{({z - L})}}cos\left( {\beta z - \epsilon} \right)cos\left( {\kappa z} \right)}{cos\left( {\beta L - \epsilon} \right)cos\left( {\kappa L} \right)}\left\lbrack {Q_{s},P_{s}} \right\rbrack = \frac{i}{2}. \label{eq:commutation relations_qsps}
\end{align}
\end{subequations}
The maintenance of the correct commutation relations (\ref{eq:commutation relations_qipi}) and (\ref{eq:commutation relations_qsps}) serves as confirmation for the accuracy of Eq.~(2) presented in the main text. In other words, the Heisenbug equation for $a_i$ should not incorporate a Langevin noise to uphold the commutation relations; otherwise, they would be disrupted.

\subsection{Inseparability of real signal and Langevin noise in measurement}
The solutions (4a) and (4b) from the main text clearly indicate that the four output quadratures are affected by Langevin noise. Consequently, the measured data comprises both real signals and inseparable Langevin noise. In the subsequent analysis, we highlight these noise contributions in red while deriving the four quadrature variances (5a)-(5b) as presented in the main text:
\begin{subequations}
\begin{align}
\left\langle {{\Delta}q^2_i(0)} \right\rangle &= \left\langle {q_{i}^{2}(0)} \right\rangle - \left\langle {q_{i}(0)} \right\rangle^{2}\nonumber\\&={sec}^{2}\left( {\beta L - \epsilon} \right){\cos^{2}\epsilon}\left\langle {q_{i}^{2}(L)} \right\rangle + {sec}^{2}\left( {\beta L - \epsilon} \right){\sin^{2}\left( {\beta L} \right)}\left\langle {p_{s}^{2}(0)} \right\rangle\nonumber\\&+ {sec}^{2}\left( {\beta L - \epsilon} \right)\left\langle {\int_{0}^{L}{P_{s}(z){\sin\left( {\beta L - \beta z} \right)}{\int_{0}^{L}{P_{s}\left( z^{'} \right){\sin\left( {\beta L - \beta z^{'}} \right)}dz^{'}}}dz}} \right\rangle\nonumber\\ &= \frac{1}{4}{sec}^{2}\left( {\beta L - \epsilon} \right){\cos^{2}\epsilon} + \frac{1}{4}{sec}^{2}\left( {\beta L - \epsilon} \right){\sin^{2}\left( {\beta L} \right)}\nonumber\\&+ \frac{\gamma}{2}{sec}^{2}\left( {\beta L - \epsilon} \right){\int_{0}^{L}{{\sin\left( {\beta L - \beta z} \right)}{\int_{0}^{L}{\delta\left( {z - z^{'}} \right){\sin\left( {\beta L - \beta z^{'}} \right)}dz^{'}}}dz}} \nonumber\\&= \frac{1}{4}{sec}^{2}\left( {\beta L - \epsilon} \right){\cos^{2}\epsilon} + \frac{1}{4}{sec}^{2}\left( {\beta L - \epsilon} \right){\sin^{2}\left( {\beta L} \right)} \nonumber\\&+ \frac{\gamma}{2}{sec}^{2}\left( {\beta L - \epsilon} \right){\int_{0}^{L}{{\sin^{2}\left( {\beta L - \beta z} \right)}dz}} \nonumber\\&= \frac{1}{4}{sec}^{2}\left( {\beta L - \epsilon} \right){\cos^{2}\epsilon} + \frac{1}{4}{sec}^{2}\left( {\beta L - \epsilon} \right){\sin^{2}\left( {\beta L} \right)} \nonumber\\&{\color{red}- \frac{\gamma}{8\beta}\left\lbrack {{\sin\left( {2\beta L} \right)} - 2\beta L} \right\rbrack{sec}^{2}\left( {\beta L - \epsilon} \right)} \nonumber\\&= \frac{2\gamma L + 2{\cos^{2}\epsilon} + 2{\sin^{2}\left( {\beta L} \right)} - {\tan\epsilon}{\sin\left( {2\beta L} \right)}}{8{cos}^{2}\left( {\beta L - \epsilon} \right)}  \nonumber\\&= \frac{3 + 2\gamma L - 2{\sin^{2}\epsilon} - \left\lbrack {{\cos(2\beta L)} + {\tan\epsilon}{\sin(2\beta L)}} \right\rbrack}{8{cos}^{2}\left( {\beta L - \epsilon} \right)} \displaybreak[1]\nonumber\\&= \frac{3 + 2\gamma L - 2{\sin^{2}\epsilon} - {\sec\epsilon}{\cos\left( {2\beta L - \epsilon} \right)}}{8{cos}^{2}\left( {\beta L - \epsilon} \right)}, \label{eq:qivariance}\\
\left\langle {{\Delta}p^2_s(L)} \right\rangle &= \left\langle {p_{s}^{2}(L)} \right\rangle - \left\langle {p_{s}(L)} \right\rangle^{2} \nonumber\\&= {sec}^{2}\left( {\beta L - \epsilon} \right){\cos^{2}\epsilon}\left\langle {p_{s}^{2}(0)} \right\rangle + {sec}^{2}\left( {\beta L - \epsilon} \right){\sin^{2}(\beta L)}\left\langle {q_{i}^{2}(L)} \right\rangle \nonumber\\&+ {sec}^{2}\left( {\beta L - \epsilon} \right)\left\langle {\int_{0}^{L}{P_{s}(z){\cos\left( {\beta z - \epsilon} \right)}{\int_{0}^{L}{P_{s}\left( z^{'} \right){\cos\left( {\beta z^{'} - \epsilon} \right)}dz^{'}}}dz}} \right\rangle \nonumber\\&= \frac{1}{4}{sec}^{2}\left( {\beta L - \epsilon} \right){\cos^{2}\epsilon} + \frac{1}{4}{sec}^{2}\left( {\beta L - \epsilon} \right){\sin^{2}(\beta L)} \nonumber\\&+ \frac{\gamma}{2}{sec}^{2}\left( {\beta L - \epsilon} \right){\int_{0}^{L}{{\cos\left( {\beta z - \epsilon} \right)}{\int_{0}^{L}{\delta\left( {z - z^{'}} \right){\cos\left( {\beta z^{'} - \epsilon} \right)}dz^{'}}}dz}} \nonumber\\&= \frac{1}{4}{sec}^{2}\left( {\beta L - \epsilon} \right){\cos^{2}\epsilon} + \frac{1}{4}{sec}^{2}\left( {\beta L - \epsilon} \right){\sin^{2}(\beta L)} \nonumber\\&+ \frac{\gamma}{2}{sec}^{2}\left( {\beta L - \epsilon} \right){\int_{0}^{L}{{\cos^{2}\left( {\beta z - \epsilon} \right)}dz}} \nonumber\\&= \frac{1}{4}{sec}^{2}\left( {\beta L - \epsilon} \right){\cos^{2}\epsilon} + \frac{1}{4}{sec}^{2}\left( {\beta L - \epsilon} \right){\sin^{2}(\beta L)} \nonumber\\&{\color{red}+ \frac{\gamma}{8\beta}\left\lbrack {{\sin\left( {2\beta L - 2\epsilon} \right)} + 2\beta L + {\sin(2\epsilon)}} \right\rbrack{sec}^{2}\left( {\beta L - \epsilon} \right) \nonumber}\\&= \frac{2\gamma L + 2{\cos^{2}\epsilon} + 2{\sin^{2}(\beta L)} + {\tan\epsilon}{\sin\left( {2\beta L - 2\epsilon} \right)} + {\tan\epsilon}{\sin(2\epsilon)}}{8{cos}^{2}\left( {\beta L - \epsilon} \right)} \nonumber\\&= \frac{2\gamma L + 2 - 2{\sin^{2}\epsilon} + 1 - {\cos(2\beta L)} + {\tan\epsilon}{\sin\left( {2\beta L - 2\epsilon} \right)} + 2{\sin^{2}\epsilon}}{8{cos}^{2}\left( {\beta L - \epsilon} \right)} \nonumber\\&= \frac{3 + 2\gamma L - {\cos(2\beta L)} + {\tan\epsilon}{\sin\left( {2\beta L - 2\epsilon} \right)}}{8{cos}^{2}\left( {\beta L - \epsilon} \right)}, \label{eq:psvariance}\displaybreak[1]\\
\left\langle {{\Delta}q^2_s(L)} \right\rangle &= \left\langle {q_{s}^{2}(L)} \right\rangle - \left\langle {q_{s}(L)} \right\rangle^{2} \nonumber\\&= e^{- 2\gamma L}{sec}^{2}\left( {\kappa L} \right)\left\langle {q_{s}^{2}(0)} \right\rangle + {tan}^{2}\left( {\kappa L} \right)\left\langle {p_{i}^{2}(L)} \right\rangle \nonumber\\&+ {sec}^{2}\left( {\kappa L} \right)\left\langle {\int_{0}^{L}{Q_{s}(z)e^{\gamma{({z - L})}}{\cos\left( {\kappa z} \right)}{\int_{0}^{L}{Q_{s}\left( z^{'} \right)e^{\gamma{({z^{'} - L})}}{\cos\left( {\kappa z^{'}} \right)}dz^{'}}}dz}} \right\rangle \nonumber\\&= \frac{1}{4}e^{- 2\gamma L}{sec}^{2}(\kappa L) + \frac{1}{4}{tan}^{2}(\kappa L) \nonumber\\&+ \frac{\gamma}{2}{sec}^{2}\left( {\kappa L} \right){\int_{0}^{L}{e^{\gamma{({z - L})}}{\cos\left( {\kappa z} \right)}{\int_{0}^{L}{\delta\left( {z - z^{'}} \right)e^{\gamma{({z^{'} - L})}}{\cos\left( {\kappa z^{'}} \right)}dz^{'}}}dz}} \nonumber\\&= \frac{1}{4}e^{- 2\gamma L}{sec}^{2}\left( {\kappa L} \right) + \frac{1}{4}{tan}^{2}\left( {\kappa L} \right)+ \frac{\gamma}{2}{sec}^{2}\left( {\kappa L} \right){\int_{0}^{L}{e^{2\gamma{({z - L})}}{\cos^{2}\left( {\kappa z} \right)}dz}} \nonumber\\&= \frac{1}{4}e^{- 2\gamma L}{sec}^{2}\left( {\kappa L} \right) + \frac{1}{4}{tan}^{2}\left( {\kappa L} \right) \nonumber\\&{\color{red}+ \frac{\kappa^{2} + \gamma^{2} - \left( {\kappa^{2} + 2\gamma^{2}} \right)e^{- 2\gamma L} + \gamma^{2}{\cos\left( {2\kappa L} \right)} + \kappa\gamma{\sin\left( {2\kappa L} \right)}}{8\left( {\kappa^{2} + \gamma^{2}} \right){cos}^{2}\left( {\kappa L} \right)} }\nonumber\\&= \frac{{\cos^{2}\varphi}e^{- 2\gamma L} + 2 - {\cos(2\kappa L)} + {\sin^{2}\varphi}{\cos(2\kappa L)} + {\sin\varphi}{\cos\varphi}{\sin(2\kappa L)}}{8{cos}^{2}\left( {\kappa L} \right)} \nonumber\\&= \frac{\left. {\cos^{2}\varphi}e^{- 2\gamma L} + 2 - {\cos^{2}\varphi}{\cos\left( 2\kappa L \right.} \right) + {\sin\varphi}{\cos\varphi}{\sin(2\kappa L)}}{8{cos}^{2}(\kappa L)} \nonumber\\&= \frac{2 + {\cos^{2}\varphi}e^{- 2\gamma L} - {\cos\varphi}{\cos\left( {2\kappa L + \varphi} \right)}}{8{cos}^{2}(\kappa L)}, \label{eq:qsvariance}\\
\left\langle {{\Delta}p^2_i(0)} \right\rangle &= \left\langle {p_{i}^{2}(0)} \right\rangle - \left\langle {p_{i}(0)} \right\rangle^{2} \nonumber\\&= e^{2\gamma L}{sec}^{2}(\kappa L)\left\langle {p_{i}^{2}(L)} \right\rangle + {tan}^{2}(\kappa L)\left\langle {q_{s}^{2}(0)} \right\rangle +\nonumber\\& {sec}^{2}(\kappa L)\left\langle {\int_{0}^{L}{Q_{s}(z)e^{\gamma z}{\sin\left( {\kappa L - \kappa z} \right)}{\int_{0}^{L}{Q_{s}\left( z^{'} \right)e^{\gamma z^{'}}{\sin\left( {\kappa L - \kappa z^{'}} \right)}dz^{'}}}dz}} \right\rangle \displaybreak[1]\nonumber\\&= \frac{1}{4}e^{2\gamma L}{sec}^{2}(\kappa L) + \frac{1}{4}{tan}^{2}(\kappa L) + \nonumber\\&\frac{\gamma}{2}{sec}^{2}(\kappa L){\int_{0}^{L}{e^{\gamma z}{\sin\left( {\kappa L - \kappa z} \right)}{\int_{0}^{L}{\delta\left( {z - z^{'}} \right)e^{\gamma z^{'}}{\sin\left( {\kappa L - \kappa z^{'}} \right)}dz^{'}}}dz}} \nonumber\\&= \frac{1}{4}e^{2\gamma L}{sec}^{2}(\kappa L) + \frac{1}{4}{tan}^{2}(\kappa L) \nonumber\\&+ \frac{\gamma}{2}{sec}^{2}(\kappa L){\int_{0}^{L}{e^{2\gamma z}{\sin^{2}\left( {\kappa L - \kappa z} \right)}dz}} \nonumber\\&= \frac{1}{4}e^{2\gamma L}{sec}^{2}(\kappa L) + \frac{1}{4}{tan}^{2}(\kappa L) \nonumber\\&{\color{red}+ \frac{\kappa^{2}e^{2\gamma L} + \gamma^{2}{\cos(2\kappa L)} - \kappa\gamma{\sin(2\kappa L)} - \left( {\kappa^{2} + \gamma^{2}} \right)}{8\left( {\kappa^{2} + \gamma^{2}} \right){cos}^{2}(\kappa L)}} \nonumber\\&= \frac{\left( {2 + {\cos^{2}\varphi}} \right)e^{2\gamma L} - {\cos(2\kappa L)} + {\sin^{2}\varphi}{\cos(2\kappa L)} - {\sin\varphi}{\cos\varphi}{\sin(2\kappa L)}}{8{cos}^{2}\left( {\kappa L} \right)}\nonumber\\& = \frac{\left( {2 + {\cos^{2}\varphi}} \right)e^{2\gamma L} - {\cos^{2}\varphi}{\cos(2\kappa L)} - {\sin\varphi}{\cos\varphi}{\sin(2\kappa L)}}{8{cos}^{2}(\kappa L)} \nonumber\\&= \frac{\left( {2 + {\cos^{2}\varphi}} \right)e^{2\gamma L} - {\cos\varphi}{\cos\left( {2\kappa L - \varphi} \right)}}{8{cos}^{2}(\kappa L)}.\label{eq:pivariance}
\end{align}
\end{subequations}
where $\varphi = {\arctan\left( \frac{\gamma}{\kappa} \right)}$.

When combined with the solutions (4a) and (4b), our calculations demonstrate that, within the CV framework, the utilization of PSA remains compliant with the quantum noncloning theorem and avoids Hilbert space expansion. For real signals, both the signal and idler modes are anticipated to share a joint Einstein-Podolsky-Rosen (EPR)-type state \cite{14}, with neither of them possessing an individual state. An investigation on how PT or anti-Hermiticity influences such EPR correlations will be addressed separately \cite{13}. Besides, according to our understanding, in our type-I quadrature-PT system, it is the Langevin noise that enlarges the Hilbert space and induces thermalization of the peculiar photon-number fluctuations between the two modes.

\subsection{Calculations of the NF}
In this subsection, we lay out the key steps in evaluating the NF (7) discussed in the main text:
\begin{align}
Var\left\lbrack {N_{i}(0) - N_{s}(L)} \right\rbrack &= \left\langle \left( {N_{i}(0) - N_{s}(L)} \right)^{2} \right\rangle - \left\langle {N_{i}(0) - N_{s}(L)} \right\rangle^{2} \nonumber\\&= \left\langle {{\Delta}N_{i}^{2}(0)} \right\rangle + \left\langle {{\Delta}N_{s}^{2}(L)} \right\rangle - 2\left\lbrack {\left\langle {N_{i}(0)N_{s}(L)} \right\rangle - \left\langle {N_{i}(0)} \right\rangle\left\langle {N_{s}(L)} \right\rangle} \right\rbrack. \label{eq:NF}
\end{align}
where 
\begin{subequations}
\begin{align}
\left\langle {{\Delta}N_{i}^{2}(0)} \right\rangle &= \left\langle \left\lbrack {q_{i}^{2}(0) + p_{i}^{2}(0) - \frac{1}{2}} \right\rbrack^{2} \right\rangle - \left\langle {q_{i}^{2}(0) + p_{i}^{2}(0) - \frac{1}{2}} \right\rangle^{2} \nonumber\\&= \left\langle {q_{i}^{4}(0)} \right\rangle - \left\langle {q_{i}^{2}(0)} \right\rangle^{2} + \left\langle {p_{i}^{4}(0)} \right\rangle - \left\langle {p_{i}^{2}(0)} \right\rangle^{2} \nonumber\\&+ \left\langle {q_{i}^{2}(0)p_{i}^{2}(0) + p_{i}^{2}(0)q_{i}^{2}(0)} \right\rangle - 2\left\langle {q_{i}^{2}(0)} \right\rangle\left\langle {p_{i}^{2}(0)} \right\rangle, \label{eq:Nivariance} \\
\left\langle {{\Delta}N_{s}^{2}(L)} \right\rangle &= \left\langle \left\lbrack {q_{s}^{2}(L) + p_{s}^{2}(L) - \frac{1}{2}} \right\rbrack^{2} \right\rangle - \left\langle {q_{s}^{2}(L) + p_{s}^{2}(L) - \frac{1}{2}} \right\rangle^{2} \nonumber\\&= \left\langle {q_{s}^{4}(L)} \right\rangle - \left\langle {q_{s}^{2}(L)} \right\rangle^{2} + \left\langle {p_{s}^{4}(L)} \right\rangle - \left\langle {p_{s}^{4}(0)} \right\rangle^{2} \nonumber\\&+ \left\langle {q_{s}^{2}(L)p_{s}^{2}(L) + p_{S}^{2}(L)q_{s}^{2}(L)} \right\rangle - 2\left\langle {q_{s}^{2}(L)} \right\rangle\left\langle {p_{s}^{2}(L)} \right\rangle, \label{eq:Nsvariance}\\
\left\langle {N_{i}(0)N_{s}(L)} \right\rangle - \left\langle {N_{i}(0)} \right\rangle\left\langle {N_{s}(L)} \right\rangle &= \left\langle {\left( {q_{i}^{2}(0) + p_{i}^{2}(0) - \frac{1}{2}} \right)\left( {q_{s}^{2}(L) + p_{s}^{2}(L) - \frac{1}{2}} \right)} \right\rangle \nonumber\\&- \left\langle {q_{i}^{2}(0) + p_{i}^{2}(0) - \frac{1}{2}} \right\rangle\left\langle {q_{s}^{2}(L) + p_{s}^{2}(L) - \frac{1}{2}} \right\rangle \nonumber\\&= \left\langle {q_{i}^{2}(0)q_{s}^{2}(L)} \right\rangle - \left\langle {q_{i}^{2}(0)} \right\rangle\left\langle {q_{s}^{2}(L)} \right\rangle \nonumber\\&+ \left\langle {q_{i}^{2}(0)p_{s}^{2}(L)} \right\rangle - \left\langle {q_{i}^{2}(0)} \right\rangle\left\langle {p_{s}^{2}(L)} \right\rangle \displaybreak[1]\nonumber\\&+ \left\langle {p_{i}^{2}(0)q_{s}^{2}(L)} \right\rangle - \left\langle {p_{i}^{2}(0)} \right\rangle\left\langle {q_{s}^{2}(L)} \right\rangle \nonumber\\&+ \left\langle {p_{i}^{2}(0)p_{s}^{2}(L)} \right\rangle - \left\langle {p_{i}^{2}(0)} \right\rangle\left\langle {p_{s}^{2}(L)} \right\rangle. \label{eq:NiNsvariance}
\end{align}
\end{subequations}
The remaining calculations involve computing the expectation values in the above formulas term by term. However, due to the lengthy expressions, we will refrain from presenting these results here. Note that for the initial vacuum input state, we have $\left\langle {q_{i}(0)} \right\rangle = \left\langle {p_{s}(L)} \right\rangle = \left\langle {p_{i}(0)} \right\rangle = \left\langle {q_{s}(L)} \right\rangle = 0$.

\begin{figure}[htbp]
\centering
\fbox{\includegraphics[width=.6\linewidth]{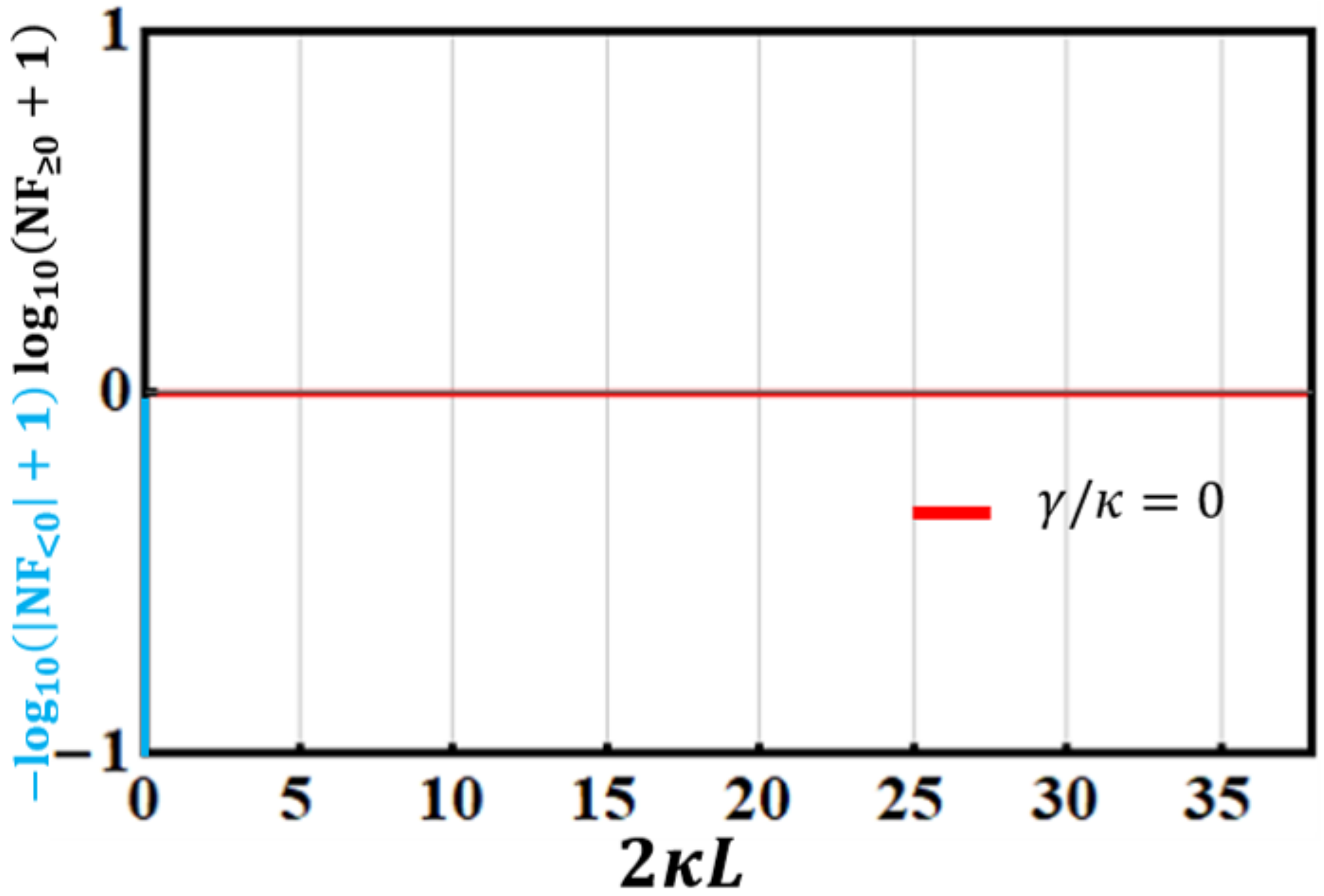}}
\caption{Examining NF in the Hermitian scenario ($g=\gamma=0,\kappa=1$) using an initial vacuum input state, to assess the efficacy of the chosen logarithmic representation in discriminating relative-intensity classical noise from relative-intensity squeezed noise.}
\label{fig:FigS1}
\end{figure}

Before proceeding to the next subsection, here we would like to provide an explanation for the format in which we plot Figs. 4(a) and (b) in the main text, using $lg\left( {NF}_{\geq 0} + 1 \right)$ and $- lg\left( \left| {NF}_{< 0} \right| + 1 \right)$. The reason for this choice is to accurately represent the logarithmic form of NF and reference NF in the Hermitian case ($g = \gamma = 0,\kappa = 1$) within a unified figure for an initial vacuum input state (Fig. ~\ref{fig:FigS1}).

\subsection{Proof of the impossibility of PIA to form quadrature PT symmetry}
If the PSA were to be replaced with PIA, the NH Hamiltonian would be represented as follows:
\begin{eqnarray}
H_{PIA} = i\hslash ga_{i}^{\dagger}a_{i} - i\hslash\gamma a_{s}^{\dagger}a_{s} + \hslash\kappa\left( a_{i}^{\dagger}a_{s}^{\dagger}e^{- i\xi} + a_{i}a_{s}e^{i\xi} \right).\label{eq:H_PIA}
\end{eqnarray}
where $\xi$ denotes the relative phase of the two input pump lasers. To demonstrate that the Hamiltonian (\ref{eq:H_PIA}) does not lead to quadrature PT (or anti-PT) symmetry, we will examine two contrasting scenarios related to Fig.~1(a) in the main text: (i) both Pump 1 and Pump 2 are directed along the $+z$-axis; and (ii) Pump 1 is reoriented to propagate in the $-z$-direction instead.

For case (i), the Heisenberg-Langevin equations of the coupled signal-idler field operators are
\begin{eqnarray}
\frac{da_{i}}{dz} = ga_{i} - i\kappa e^{- i\xi}a_{s}^{\dagger} + f_{i},~~~\frac{da_{s}}{dz} = - \gamma a_{s} - i\kappa e^{- i\xi}a_{i}^{\dagger} + f_{s}. \label{eq:PIA_equation}
\end{eqnarray}
Now if $\xi=0$, we attain the following coupled-quadrature equations,
\begin{subequations}
\begin{align}
\frac{d}{dz}\begin{bmatrix}
q_{i} \\p_{s}\end{bmatrix} = \begin{bmatrix}g & {- \kappa} \\{- \kappa} & {- \gamma}\end{bmatrix}\begin{bmatrix}q_{i} \\p_{s}\end{bmatrix} + \begin{bmatrix}Q_{i} \\P_{s}\end{bmatrix}, \label{eq:xi1_0_qips} \\
\frac{d}{dz}\begin{bmatrix}p_{i} \\q_{s}\end{bmatrix} = \begin{bmatrix}g & {- \kappa} \\{- \kappa} & {- \gamma}\end{bmatrix}\begin{bmatrix}p_{i} \\q_{s}\end{bmatrix} + \begin{bmatrix}P_{i} \\Q_{s}\end{bmatrix}.\label{eq:xi1_0_piqs}
\end{align}
\end{subequations}
with $H_{({q_{i},p_{s},\xi = 0})}^{1} = \begin{bmatrix} {ig} & {-i\kappa} \\{- i\kappa} & {-i\gamma}\end{bmatrix}$ and $H_{({p_{i},q_{s},\xi = 0})}^{1} = \begin{bmatrix} {ig} & {- i\kappa} \\ {- i\kappa} & {-i\gamma}\end{bmatrix}$.
If $\xi = \pi/2$, the two sets of coupled-quadrature equations become,
\begin{subequations}
\begin{align}
\frac{d}{dz}\begin{bmatrix}q_{i} \\q_{s}\end{bmatrix} = \begin{bmatrix}g& {- \kappa} \\{- \kappa} & {- \gamma}\end{bmatrix}\begin{bmatrix}q_{i} \\q_{s}\end{bmatrix} + \begin{bmatrix}Q_{i} \\Q_{s}\end{bmatrix}, \label{eq:xi1_pi/2_qiqs}\\
\frac{d}{dz}\begin{bmatrix}p_{i} \\p_{s}\end{bmatrix} = \begin{bmatrix}g & {- \kappa} \\{- \kappa} & {- \gamma}\end{bmatrix}\begin{bmatrix}p_{i} \\p_{s}\end{bmatrix} + \begin{bmatrix}P_{i} \\P_{s}\end{bmatrix}.\label{eq:xi1_pi/2_pips}
\end{align}
\end{subequations}
with $H_{({q_{i},q_{s},\xi = \pi/2})}^{1} = \begin{bmatrix} {ig} & {-i\kappa} \\{- i\kappa} & {-i\gamma}\end{bmatrix}$ and $H_{({p_{i},p_{s},\xi = \pi/2})}^{1} = \begin{bmatrix} {ig} & {- i\kappa} \\ {- i\kappa} & {-i\gamma}\end{bmatrix}$.
One can easily check that when $g = \gamma$, none of $H_{({q_{i},p_{s},\xi = 0})}^{1}$, $H_{({p_{i},q_{s},\xi = 0})}^{1}$, $H_{({q_{i},q_{s},\xi = \pi/2})}^{1}$, and $H_{({p_{i},p_{s},\xi = \pi/2})}^{1}$ satisfies PT or anti-PT symmetry.

For case (ii), the Heisenberg-Langevin equations of the coupled signal-idler field operators are
\begin{eqnarray}
\frac{da_{i}}{dz} =- ga_{i} + i\kappa e^{- i\xi}a_{s}^{\dagger} - f_{i},~~~\frac{da_{s}}{dz} = - \gamma a_{s} - i\kappa e^{- i\xi}a_{i}^{\dagger} + f_{s}. \label{eq:PIA_equation}
\end{eqnarray}
In the similar way, when $\xi=0$, we have the following coupled-quadrature equations,
\begin{subequations}
\begin{align}
\frac{d}{dz}\begin{bmatrix}
q_{i} \\p_{s}\end{bmatrix} = \begin{bmatrix}-g & { \kappa} \\{- \kappa} & {- \gamma}\end{bmatrix}\begin{bmatrix}q_{i} \\p_{s}\end{bmatrix} + \begin{bmatrix}-Q_{i} \\P_{s}\end{bmatrix}, \label{eq:xi2_0_qips} \\
\frac{d}{dz}\begin{bmatrix}p_{i} \\q_{s}\end{bmatrix} = \begin{bmatrix}-g & {- \kappa} \\{- \kappa} & {- \gamma}\end{bmatrix}\begin{bmatrix}p_{i} \\q_{s}\end{bmatrix} + \begin{bmatrix}-P_{i} \\Q_{s}\end{bmatrix}.\label{eq:xi2_0_piqs}
\end{align}
\end{subequations}
with $H_{({q_{i},p_{s},\xi = 0})}^{2} = \begin{bmatrix} {-ig} & {i\kappa} \\{- i\kappa} & {-i\gamma}\end{bmatrix}$ and $H_{({p_{i},q_{s},\xi = 0})}^{2} = \begin{bmatrix} {-ig} & {i\kappa} \\ {- i\kappa} & {-i\gamma}\end{bmatrix}$. For $\xi=pi/2$, the two sets of coupled-quadrature equations become,
\begin{subequations}
\begin{align}
\frac{d}{dz}\begin{bmatrix}
q_{i} \\q_{s}\end{bmatrix} = \begin{bmatrix}-g & { \kappa} \\{- \kappa} & {- \gamma}\end{bmatrix}\begin{bmatrix}q_{i} \\q_{s}\end{bmatrix} + \begin{bmatrix}-Q_{i} \\Q_{s}\end{bmatrix}, \label{eq:xi2_pi/2_qiqs} \\
\frac{d}{dz}\begin{bmatrix}p_{i} \\p_{s}\end{bmatrix} = \begin{bmatrix}-g & {- \kappa} \\{ \kappa} & {- \gamma}\end{bmatrix}\begin{bmatrix}p_{i} \\q_{s}\end{bmatrix} + \begin{bmatrix}-P_{i} \\P_{s}\end{bmatrix}.\label{eq:xi2_pi/2_pips}
\end{align}
\end{subequations}
with $H_{({q_{i},q_{s},\xi = pi/2})}^{2} = \begin{bmatrix} {-ig} & {i\kappa} \\{- i\kappa} & {-i\gamma}\end{bmatrix}$ and $H_{({p_{i},p_{s},\xi = pi/2})}^{2} = \begin{bmatrix} {-ig} & {-i\kappa} \\ {i\kappa} & {-i\gamma}\end{bmatrix}$. Again, one can show that, for $g = \gamma$, neither $H_{({q_{i},p_{s},\xi = 0})}^{2}$, $H_{({p_{i},q_{s},\xi = 0})}^{2}$, $H_{({q_{i},q_{s},\xi = \pi/2})}^{2}$, nor $H_{({p_{i},p_{s},\xi = \pi/2})}^{2}$ satisfy PT or anti-PT symmetry.

\subsection{Verification of the SU(1,1) symmetry of the NH Hamiltonians}

Consider certain angular momentum operators $L_x$, $L_y$ and $L_z$ satisfying the commutation relations
\begin{eqnarray}
[L_x,L_y]=i\sigma L_z,\; [L_y,L_z]=i L_x,\; [L_z,L_x]=i L_y, \label{eq:angular_momentum_operators1}
\end{eqnarray}
Then the operators $L_z$,$L_{\pm}=L_x\pm i L_y$ commute according to 
\begin{eqnarray}
[L_z,L_{\pm}]=\pm L_{\pm},\; [L_{+},L_{-}]=2\sigma L_z. \label{eq:angular_momentum_operators2}
\end{eqnarray}
It is clear that $\sigma=1$ corresponds to the $SU(2)$-Lie algebra, whereas $\sigma=-1$ corresponds to the $SU(1,1)$-Lie algebra. For the non-Hermitian (NH) Hamiltonian $H$ (Eq. (1)) as provided in the main text, we have the Hermitian operators
\begin{subequations}
\begin{align}
L_{i,-}=\frac{1}{2}{a^2_{i}},\; L_{i,+}=\frac{1}{2}a_{i}^{\dagger^2},\; L_{i,z}=\frac{1}{2}(a_{i}^{\dagger}a_{i}+\frac{1}{2}),\label{eq:su11operators_idler}\\L_{-}=a_{i}a_{s},\; L_{+}=a^{\dagger}_{i}a^{\dagger}_{s},\; L_{z}=\frac{1}{2}(a^{\dagger}_{i}a_{i}+a^{\dagger}_{s}a_{s}+1),
\label{eq:su11operators_idler_and_signal}
\end{align}
\end{subequations}
According to the definition (\ref{eq:angular_momentum_operators2}), the operators (\ref{eq:su11operators_idler}) and (\ref{eq:su11operators_idler_and_signal}) satisfy the commutation relation relations for the Lie algebra of $SU(1,1)$. Therefore, the non-Hermitian (NH) Hamiltonian $H$ (Eq. (1)) can be rewritten as
\begin{eqnarray}
H=i\hbar g(L_{i,-}-L_{i,+})-2i\hbar\gamma(L_z-L_{i,z}-\frac{1}{4})+\hbar\kappa(L_{-}+L_{+}),\label{eq:hamiltonian_su11}
\end{eqnarray}
As one can see, the above Hamiltonian represents the interaction between two $SU(1,1)$ systems. The presence of this supplementary symmetry is anticipated to open avenues for novel effects deserving of further exploration. Our conjecture draws inspiration from our recent studies \cite{10,11}, where the PT-symmetric but NH Hamiltonian governing a dissipative spin-boson-coupled superconducting circuit system adheres to SU(1) symmetry, specifically the conservation of excitation number. This unique symmetry gives rise to exceptional entanglement transitions, a phenomenon absent in classical and Hermitian quantum counterparts.

%On the other hand, the most general complex Hamiltonian quadratic in the $SU(2)$ operators belongs to an 18-parameter family \cite{15}
%\begin{eqnarray}
%H=\sum_{j=1}^{2} C_j L_j^{'}+\sum_{0\leq j<k\leq 2} C_{jk} L_j^{'} L_k^{'},\label{eq:hamiltonian_su2}
%\end{eqnarray}
%where $C_j$ and $C_{jk}$ is complex number, $L_0^{'}\equiv L_z^{'}$, $L_1^{'}\equiv L_x^{'}$ and $L_2^{'}\equiv L_y^{'}$. 
%For the effective NH Hamiltonian matrices (\ref{eq:effect_H_qips}) and (\ref{eq:effect_H_piqs}), we find the angular momentum operators
%\begin{eqnarray}
%L_x^{'}=\frac{1}{2}\begin{bmatrix}0 & 1 \\1 & 0\end{bmatrix}, L_y^{'}=\frac{1}{2}\begin{bmatrix}0 & -i \\i & 0\end{bmatrix}, L_z^{'}=\frac{1}{2}\begin{bmatrix}1 & 0 \\0 & -1\end{bmatrix}. \label{eq:J_angular_momentum_operators}
%\end{eqnarray}
%which satisfy the commutation relation relations for the Lie algebra of $SU(2)$ (\ref{eq:angular_momentum_operators2}). Then the effective NH Hamiltonian matrices (\ref{eq:effect_H_qips}) and (\ref{eq:effect_H_piqs}) can be expressed as
%\begin{eqnarray}
%H_{({q_{i},p_{s}})} = \begin{bmatrix}
%{i\gamma} & {i\kappa} \\
%{- i\kappa} & {- i\gamma}
%\end{bmatrix}=2i\gamma L_0^{'}+4i\kappa L_0^{'} L_1^{'},\label{eq:J_qips}\\
%H_{({p_{i},q_{s}})} = \begin{bmatrix}
%{-i\gamma} & {i\kappa} \\
%{- i\kappa} & {- i\gamma}
%\end{bmatrix}=-i\gamma \mathcal{I}+4i\kappa L_0^{'} L_1^{'},\label{eq:J_qips}
%\end{eqnarray}
%which represent these dual mode quadrature effective Hamiltonian can be viewed as a $SU(2)$ system commonly used to describe atomic behavior. 

\section{Further discussion on quantum sensing}
Utilizing Eqs. (9) and (10) given in the main text, we are ready to work out $\Delta\kappa^2_{q_{i,0}}=\frac{\langle\Delta q^2_{i}(0)\rangle}{\left(\chi^{q_i(0)}_{\kappa}\right)^2}$. Its inverse thus takes the following expression:
\begin{equation}
\Delta\kappa^{-2}_{q_i(0)}\equiv\frac{\left(\chi^{q_i(0)}_{\kappa}\right)^2}{\langle\Delta q^2_i(0)\rangle}=\frac{\alpha^2\{2\beta L[\sin(\beta L-\epsilon)-1]+\sin\epsilon[2\sin(\beta L)+\cos(\beta L-\epsilon)-\cos\epsilon]\}^2}{\beta^2\cos^2(\beta L-\epsilon)\{3+w[2\gamma L-\tan\epsilon\sin(2\beta L)]-\cos(2\beta L)-2\sin^2\epsilon\}},
\end{equation}
which determines the SNR in measurement.

\subsection{Calculations of the quantum Fisher information $F_{\kappa}$}
For Gaussian states (e.g., our case), the quantum Fisher information takes the form
\begin{eqnarray}
F_{\kappa} = \left( \frac{d\mu_{out}}{d\kappa} \right)^{T}V_{out}^{- 1}\frac{d\mu_{out}}{d\kappa}. \label{eq:F_kappa}
\end{eqnarray}
where the amplitude vector $\mu_{out}$ and the variance matrix $V_{out}$ can be defined in the quadrature basis via $\mu_{j} = \langle {\hat{v}}_{j} \rangle$ and $V_{j,k} = \frac{1}{2}\langle {{\hat{v}}_{j}{\hat{v}}_{k} + {\hat{v}}_{k}{\hat{v}}_{j}} \rangle - \langle {\hat{v}}_{j} \rangle\langle {\hat{v}}_{k}\rangle$, for $1 \leq j,k \leq 2$, with the column vector $\hat{v} = \left( {q_{i}(0),q_{s}(L),p_{i}(0),p_{s}(L)} \right)^{T}$. In general, $F_{\kappa}$ (\ref{eq:F_kappa}) characterizes the SNR of the system.

For our scheme, we can show that $V_{out}$ and $\frac{d\mu_{out}}{d\kappa}$ assume the following forms,
\begin{subequations}
\begin{align}
V_{out} &= 
\begin{bmatrix}
{Var\left\lbrack {q_{i}(0)} \right\rbrack} & 0 & 0 & {CoVar\left\lbrack {q_{i}(0),p_{s}(L)} \right\rbrack}\\
0 & {Var\left\lbrack {q_{s}(L)} \right\rbrack} & {CoVar\left\lbrack {p_{i}(0),q_{s}(L)} \right\rbrack} & 0\\
0 & {CoVar\left\lbrack {p_{i}(0),q_{s}(L)} \right\rbrack} & {Var\left\lbrack {p_{i}(0)} \right\rbrack} & 0\\
{CoVar\left\lbrack {q_{i}(0),p_{s}(L)} \right\rbrack} & 0 & 0 & {Var\left\lbrack {p_{s}(L)} \right\rbrack}
\end{bmatrix}, \label{eq:Vout}\\
\frac{d\mu_{out}}{d\kappa} &= \left\lbrack {\chi_{\kappa}^{q_{i}{(0)}},\chi_{\kappa}^{q_{s}{(L)}},\chi_{\kappa}^{p_{i}{(0)}},\chi_{\kappa}^{p_{s}{(L)}}} \right\rbrack^{T}, \label{eq:muout}
\end{align}
\end{subequations}
respectively. In Eq.~(\ref{eq:Vout}), we have defined the following notions, $Var[x]=\langle\Delta x^2\rangle$ and $CoVar[x,y]=\langle xy\rangle-\langle x\rangle\langle y\rangle$, to simplify the expression. The comprehensive formulation of the quantum Fisher information is highly intricate and lacks intuitiveness. Therefore, we refrain from presenting its explicit expression in this context. Instead, we focus on providing the simplified expression when $\gamma\rightarrow\kappa$, which is
\begin{eqnarray}
\begin{aligned}
&F_{\kappa} = \frac{4}{3}\alpha^{2}L^{2}6{\sec^{3}(\kappa L})\left\{ {\frac{2\left( {\kappa^{2}L^{3} + 3L} \right)^{2}}{6{\sec^{3}(\kappa L)}\left( {\kappa L + 1} \right)\left( {\kappa^{4}L^{4} + 4\kappa^{3}L^{3} - 3\kappa^{2}L^{2} + 3} \right)}
}\right.\\&
\left.{+ \frac{\left( {\sin{\kappa L}} - {{e}}^{\kappa L} \right)\left\lbrack {{{e}}^{\kappa L}{\sin{\kappa L}} + {{e}}^{3\kappa L}{\sin{\kappa L}} - {\sin{2\kappa L}} + \left( {{{e}}^{\kappa L} - 5{{e}}^{3\kappa L}} \right){\cos{\kappa L}} + {\cos{2\kappa L}}- {{e}}^{2\kappa L}} \right\rbrack}{7{{e}}^{4\kappa L} + 2{{e}}^{2\kappa L}{\cos{2\kappa L}} - 1}}\right.\\&\left.{- \frac{\left( {{e}}^{\kappa L}{\sin{\kappa L}} - 1 \right)\left\lbrack {{{e}}^{2\kappa L}{\sin{\kappa L}} + {\sin{\kappa L}} + {{e}}^{3\kappa L}{\sin{2\kappa L}} + \left( {3{{e}}^{2\kappa L} + 1} \right){\cos{\kappa L}} + {{e}}^{3\kappa L}{\cos{2\kappa L}}- {{e}}^{\kappa L}} \right\rbrack}{7{{e}}^{4\kappa L} + 2{{e}}^{2\kappa L}{\cos{2\kappa L}} - 1}}
\right\}.
\end{aligned}
\end{eqnarray}

\subsection{Calculations of the other three susceptibilities}
If taking into account the thermal reservoir with an average thermal bosonic number $n_{th} = \left\lbrack {e^{\left( \frac{hv_{\lambda}}{k_{B}T} \right)} - 1} \right\rbrack^{- 1}$, the quantum Langevin noise in Eq.~(2) of the main text obeys the following properties, $\left\langle {f_{s}(z)f_{s}^{\dagger}\left( z^{'} \right)} \right\rangle = 2\gamma\left( {n_{th} + 1} \right)\delta\left( {z - z^{'}} \right)$ and $\left\langle {f_{s}^{\dagger}(z)f_{s}\left( z^{'} \right)} \right\rangle = 2\gamma n_{th}\delta\left( {z - z^{'}} \right)$. However, in the visible spectral range, the thermal photon number becomes infinitesimal at the room temperature ($\sim300~K$). As such, throughout this work we assume $n_{th} = 0$ to simplify the calculations.

Following the definition of the susceptibility (10) introduced in the main text, we can readily derive the system responses, i.e., the susceptibilities, to the perturbation on $\kappa$ based on measuring the other three quadratures, $q_{s}(L)$, $p_{s}(L)$, and $p_{i}(0)$. For the seeding two-photon coherent state $\left|\left.\psi_{i}\right\rangle\right.=\left|\left.{\alpha,\alpha}\right\rangle\right.$, the mean or expectation values of these three quadratures are computed to be
\begin{subequations}
\begin{align}
\left\langle {p_{s}(L)} \right\rangle &= \frac{- {\sin\left( {\beta L} \right)}}{cos\left( {\beta L - \epsilon} \right)}\left\langle {q_{i}(L)} \right\rangle + \frac{\cos\epsilon}{cos\left( {\beta L - \epsilon} \right)}\left\langle {p_{s}(0)} \right\rangle, \label{eq:ps_expectedvalue}\\
\left\langle {q_{s}(L)} \right\rangle &=-\frac{sin(\kappa L)}{cos\left( {\kappa L} \right)}\left\langle {p_{i}(L)} \right\rangle + \frac{e^{- \gamma L}}{cos\left( {\kappa L} \right)}\left\langle {q_{s}(0)} \right\rangle, \label{eq:qs_expectedvalue}\\
\left\langle {p_{i}(0)} \right\rangle &= \frac{e^{\gamma L}}{cos\left( {\kappa L} \right)}\left\langle {p_{i}(L)} \right\rangle - \frac{sin(\kappa L)}{cos\left( {\kappa L} \right)}\left\langle {q_{s}(0)} \right\rangle, \label{eq:pi_expectedvalue} 
\end{align}
\end{subequations}
and the corresponding susceptibilities are
\begin{subequations}
\begin{align}
\chi_{\kappa}^{p_{s}{(L)}} &= \frac{\partial\left\langle {p_{s}(L)} \right\rangle}{\partial\kappa} = ~\chi_{\kappa}^{q_{i}{(0)}} \nonumber\\&= \frac{\alpha\left\{ {2\beta L\left\lbrack {sin\left( {\beta L} \right) - 1} \right\rbrack + {\sin\epsilon}\left\lbrack {2sin\left( {\beta L} \right) + {\cos\left( {\beta L - \epsilon} \right)} - cos\epsilon} \right\rbrack} \right\}}{2\beta{cos}^{2}\left( {\beta L - \epsilon} \right)}, \label{eq:psxk}\\
\chi_{\kappa}^{q_{s}(L)} &= \frac{\partial\left\langle {q_{s}(L)} \right\rangle}{\partial\kappa} = \alpha L~{sec}^{2}(\kappa L)\left\lbrack {{\mathbb{e}}^{- \gamma L}sin(\kappa L) - 1} \right\rbrack, \label{eq:qsxk}\\
\chi_{\kappa}^{p_{i}(0)} &= \frac{\partial\left\langle {p_{i}(0)} \right\rangle}{\partial\kappa} = \alpha L{~sec}^{2}(\kappa L)\left\lbrack {{\mathbb{e}}^{\gamma L}sin(\kappa L) - 1} \right\rbrack, \label{eq:pixk} 
\end{align}
\end{subequations}

\begin{figure}[htbp]
\centering
\fbox{\includegraphics[width=.6\linewidth]{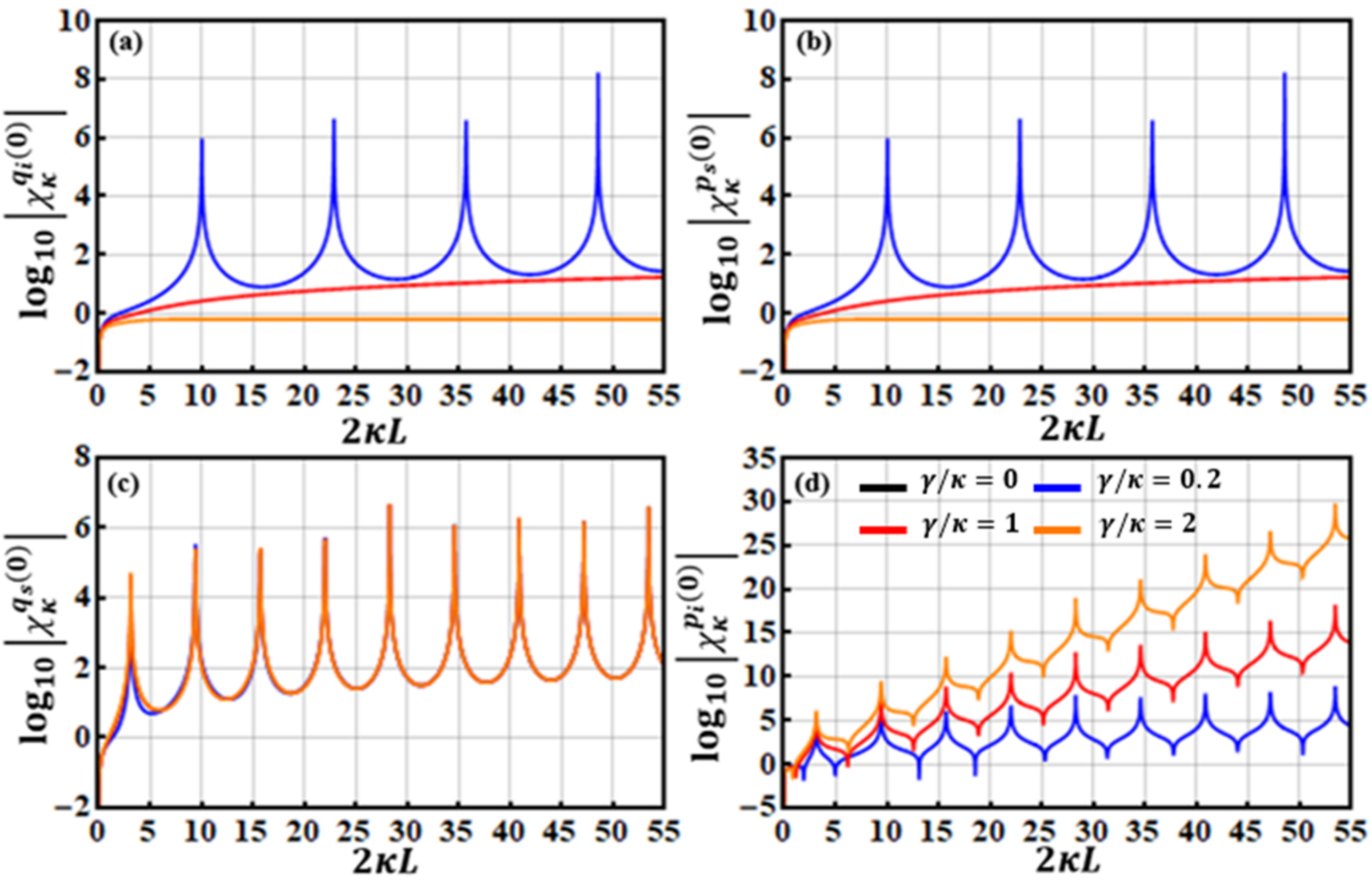}}
\caption{Exemplary susceptibilities or system responses to perturbations on $\kappa$ across different $\gamma/\kappa$ ratios, depicted as a function of $2\kappa L$.}
\label{fig:FigS2}
\end{figure}

In Fig.~\ref{fig:FigS2}, we present representative examples of the four susceptibilities for various $\gamma/{\kappa}$ values. As one can see in Figs. ~\ref{fig:FigS2}(a) and (b), in the phase intact region ($\gamma/{\kappa = 0.2}$), the PT quadrature pair $\left\{ {q_{i}(0),p_{s}(L)} \right\}$ exhibits sharp responses around the peak locations of $2\kappa L = 2\pi\kappa/\beta$. However, when the symmetry spontaneous breaks down ($\gamma/{\kappa = 1.2}$), they both lose their curvature nature and become exponential curves, indicating significantly reduced sensitivity of the system to perturbations of the measured quantities $\kappa$. In contrast, the non-PT-symmetric quadrature pair $\left\{ {q_{s}(L),p_{i}(0)} \right\}$ shown in Figs.~\ref{fig:FigS2}(c) and (d) consistently demonstrates rapid responses around the peak locations of $2\kappa L = 2\pi\left( n - \frac{1}{2} \right)$ with $n$ being positive integers, regardless of the existence of PT symmetry.

\subsection{Calculations of the other three inverse variances ${{\Delta}{\kappa}}_{{q}_{{j}}}^{- 2}$ and ${{\Delta}{\kappa}}_{{p}_{{j}}}^{-2}$}
\begin{figure}[htbp]
\centering
\fbox{\includegraphics[width=.6\linewidth]{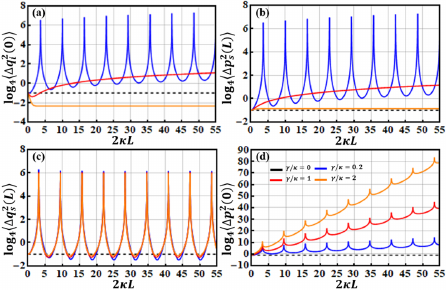}}
\caption{Both C2Q transitions in noise fluctuations and quadrature squeezing induced by loss are also evident in the case involving an input two-photon coherent state.}
\label{fig:FigS3}
\end{figure}

We have conducted similar calculations as presented in the main text to derive the other three inverse variances: ${{\Delta}\kappa}_{p_s(L)}^{- 2}$, ${{\Delta}\kappa}_{q_s(L)}^{- 2}$, and ${{\Delta}\kappa}_{p_i(0)}^{- 2}$. To obtain their expressions, one must first obtain the corresponding quadrature variances for the seeding two-photon coherent state $\left| \left. \psi_{i} \right\rangle \right. = \left| \left. {\alpha,\alpha} \right\rangle \right.$. After some algebraic manipulation, we arrive at the following results:
\begin{subequations}
\begin{align}
\langle\Delta p^2_s(L)\rangle&=\frac{3+2\gamma L-\cos(2\beta L)+\tan\epsilon\sin(2\beta L-2\epsilon)}{8\cos^2(\beta L-\epsilon)}, \label{eq:psvar}\\
\langle\Delta q^2_s(L)\rangle&=\frac{2+\cos^2\varphi e^{-2\gamma L}-\cos\varphi\cos(2\kappa L+\varphi)}{8\cos^2(\kappa L)}, \label{eq:qsvar}\\
\langle\Delta p^2_i(0)\rangle&=\frac{(2+\cos^2\varphi)e^{2\gamma L}-\cos\varphi\cos(2\kappa L-\varphi)}{8\cos^2(\kappa L)}, \label{eq:pivar} 
\end{align}
\end{subequations}
which are essentially the same as Eqs. (5b)-(5d) in the main text. As a consequence, we expect these variances to display a similar behavior to those plotted in Fig. 2 in the main text. In Fig.~\ref{fig:FigS3}, we have plotted these variances for three different $\gamma/\kappa$ values, 0.2, 1, and 2, for comparison. Although the two scenarios, one with vacuum input state and the other with two-photon coherent input state, share the same variances, the displayed physics from this latter case is significant as it also showcases the dynamical and stationary C2Q transitions and the anomalous loss-induced quadrature squeezing.

The ultimate precision of the parameter estimation of $\kappa$ is essentially determined by the inverse variance ${\Delta\kappa}^{- 2}$ associated with the physical observable of interest. The performance of the proposed sensing scheme can be however evaluated by comparing ${{\Delta}\kappa}_{q_{s}(L)}^{- 2} = \frac{\left( \chi_{\kappa}^{q_{s}(L)} \right)^{2}}{\left\langle {{\Delta}q_{s}^{2}(L)} \right\rangle}$, ${{\Delta}\kappa}_{p_{s}(L)}^{- 2} = \frac{\left( \chi_{\kappa}^{p_{s}(L)} \right)^{2}}{\left\langle {{\Delta}p_{s}^{2}(L)} \right\rangle}$, and ${{\Delta}\kappa}_{p_{i}(0)}^{- 2} = \frac{\left( \chi_{\kappa}^{p_{i}(0)} \right)^{2}}{\left\langle {{\Delta}p_{i}^{2}(0)} \right\rangle}$  with the quantum Fisher information $F_{\kappa}$ (\ref{eq:F_kappa}) at the system’s final state, as $F_{\kappa} \geq \left\{ {{{\Delta}\kappa}_{q_{s}(L)}^{- 2},{{\Delta}\kappa}_{p_{s}(L)}^{- 2},{{\Delta}\kappa}_{p_{i}(0)}^{- 2}} \right\}$. With the help of Eqs.(\ref{eq:psxk})-(\ref{eq:pixk}) and Eqs.(\ref{eq:psvar})-(\ref{eq:pivar}), we reach the following key results:
\begin{subequations}
\begin{align}
{{\Delta}\kappa}_{p_{s}(L)}^{- 2} &= \frac{2\alpha^{2}\left\{ {2\beta L\left\lbrack {{\sin(\beta L)} - 1} \right\rbrack + {\sin\epsilon}\left\lbrack {2{\sin(\beta L)} + {\cos\left( {\beta L - \epsilon} \right)} - {\cos\epsilon}} \right\rbrack} \right\}^{2}}{\beta^{2}{cos}^{2}\left( {\beta L - \epsilon} \right)\left\{ {3 + \left\lbrack {2\gamma L + {\tan\epsilon}{\sin\left( {2\beta L - 2\epsilon} \right)}} \right\rbrack - {\cos(2\beta L)}} \right\}}, \label{eq:psdk}\\
{{\Delta}\kappa}_{q_{s}(L)}^{- 2} &= \frac{8\alpha^{2}L^{2}{{sec}^{2}(\kappa L)\left\lbrack {{\mathbb{e}}^{- \gamma L}{\sin(\kappa L)} - 1} \right\rbrack}^{2}}{2 - {\cos\varphi}{\cos\left( {2\kappa L + \varphi} \right)} + {cos}^{2}\varphi{e}^{- 2\gamma L}}, \label{eq:qsdk}\\
{{\Delta}\kappa}_{p_{i}(0)}^{- 2} &= \frac{8\alpha^{2}L^{2}{{sec}^{2}(\kappa L)\left\lbrack {{\mathbb{e}}^{\gamma L}{\sin(\kappa L)} - 1} \right\rbrack}^{2}}{\left( {2 + {\cos^{2}\varphi}} \right)e^{2\gamma L} - {\cos\varphi}{\cos\left( {2\kappa L - \varphi} \right)}}. \label{eq:pidk} 
\end{align}
\end{subequations}

\begin{figure}[htbp]
\centering
\fbox{\includegraphics[width=.6\linewidth]{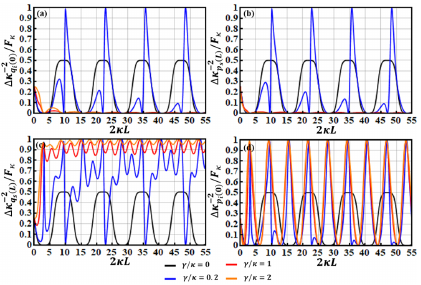}}
\caption{In relation to Fig. 5 in the main text, an alternative approach to contrasting inverse variances of four single-mode quadratures and the quantum Fisher information involves plotting their ratios.}
\label{fig:FigS4}
\end{figure}

In Figs.~5(c)-5(d) of the main text, we have illustrated these three variances in comparison with $F_{\kappa}$. Indeed, the inequality of $F_{\kappa} \geq \left\{ {{{{\Delta}\kappa}_{q_{i}(0)}^{- 2},{{\Delta}\kappa}_{p_{s}(L)}^{- 2},{\Delta}\kappa}_{q_{s}(L)}^{- 2},{{\Delta}\kappa}_{p_{i}(0)}^{- 2}} \right\}$ is well satisfied. To examine this further, we have plotted their ratios in Fig.~\ref{fig:FigS4}. This provides a clear picture of how the inverse variances behave with respect to the quantum Fisher information. With the help of Figs.~\ref{fig:FigS2} and ~\ref{fig:FigS3}, we find that for the PT-quadrature pair, optimal classical sensing is available in the PT-phase unbroken region, where $2\kappa L = 2\pi\kappa\left( 2n - \frac{1}{2} \right)/\beta$, with n being positive integers. In contrast, for the non-PT quadrature pair, both optimal classical and quantum sensing is achievable. Specifically, when $2\kappa L = 2\pi\left( 2n - \frac{1}{2} \right)$, optimal classical sensing can be achieved using either $q_{s}(L)$ or $p_{i}(0)$. However, when $2\kappa L = 2n\pi$, quantum sensing is realizable using $q_{s}(L)$ due to the occurrence of quantum squeezing. Moreover, the larger the $\gamma/\kappa$ values, the better sensitivity and SNR seem to be. While type-I quadrature PT symmetry appears to offer no advantage in enhancing quantum sensitivity, our ongoing project on type-II quadrature PT symmetry \cite{13} does demonstrate significant PT-enhanced quantum sensitivity within the PT phase unbroken region, albeit away from the Exceptional Point (EP).

\begin{figure}[htbp]
\centering
\fbox{\includegraphics[width=.6\linewidth]{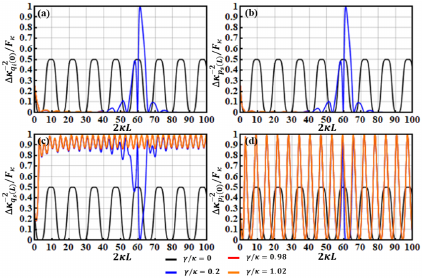}}
\caption{Evaluating sensing performance in the vicinity of the EP ($\frac{\gamma}{\kappa} = 1$) by contrasting inverse variances of four single-mode quadratures with quantum Fisher information for $\frac{\gamma}{\kappa} =$ 0.98 (blue) and 1.02 (orange).}
\label{fig:FigS5}
\end{figure}

What transpire when quantum sensing is attempted in proximity to the EP? In such circumstances, the sensing performance mirrors the patterns depicted in Fig.~\ref{fig:FigS4} above. To illustrate this point, Fig.~\ref{fig:FigS5} presents two such examples with $\gamma/\kappa$ values of 0.98 (indicated by blue lines) and 1.02 (depicted by orange lines). Evidently, prior to the spontaneous breakdown of PT symmetry, the PT-quadrature pair offers optimal classical sensing capabilities at the turning points of the recurring oscillations. Conversely, for the non-PT-symmetric quadrature pair, $q_{s}(L)$ can facilitate optimal quantum sensing with both enhanced sensitivity and SNR, whereas $p_{i}(0)$ exclusively provides optimal classical sensing performance at the inflection points of the periodic oscillations. 

On the other hand, in two recent discrete-variable experiments involving dissipative spin-boson-coupled superconducting circuit systems, the joint enhancement of quantum sensitivity through quantum entanglement and PT symmetry \cite{11} or Exceptional Points (EP) \cite{12} has been distinctly demonstrated by postselecting measurement data. In conjunction with the aforementioned conclusions, this leads us to conjecture that a universal determination regarding the feasibility of EP-enhanced sensitivity in the quantum setting might be unattainable without specifying the system details and the adopted measurement approach.

